\documentclass[oneside,reqno,10pt]{amsart}

\usepackage{amsmath,amsfonts,amssymb,mathrsfs,array,
    stmaryrd,indentfirst,amsthm,comment,color,mathtools,bbm}
\usepackage{graphicx,hyperref,enumitem,tabularx}
\usepackage{verbatim}

\usepackage{geometry}
\usepackage{natbib}
\usepackage{subfig}

\numberwithin{equation}{section}
\theoremstyle{plain}
\newtheorem{lemma}{Lemma}[section]
\newtheorem{theorem}{Theorem}[section]
\newtheorem{definition}{Definition}[section]
\newtheorem{proposition}{Proposition}[section]

\newtheorem{assumption}{Assumption}[section]

\theoremstyle{remark}
\newtheorem{remark}{Remark}[section]

\newcommand{\ignore}[1]{}

\newcommand{\cE}{\mathcal{E}}
\newcommand{\cF}{\mathcal{F}}

\newcommand{\cP}{\mathcal{P}}

\newcommand{\cS}{\mathcal{S}}

\def \a{\alpha}
\def \e{\varepsilon}

\def \1{\mathbf 1}

\def\E{\mathbb{E}}

\def\N{\mathbb{N}}
\def\P{\mathbb{P}}

\def\R{\mathbb{R}}
\def\Z{\mathbb{Z}}

\newcommand{\comm}[1]{}

\makeatletter
\def\@setcopyright{}
\def\serieslogo@{}
\makeatother

\usepackage{etoolbox, xargs, mathtools}

\usepackage{stmaryrd}
\usepackage[T1]{fontenc}
\usepackage{libertine}
\usepackage{bm} 

\usepackage[usenames,dvipsnames]{xcolor}
\usepackage{hyperref}
\hypersetup{ colorlinks, citecolor=Violet, linkcolor=NavyBlue, urlcolor=Blue}

\setlist{topsep=0.0ex, parsep=0.5ex}

\newcommand{\define}[1]{{\textbf{#1}}}

\providecommand{\alias}{}
\renewcommand{\alias}[1]{\providecommand{#1}{}\renewcommand{#1}}

\DeclarePairedDelimiter\ab{\langle}{\rangle} 
\DeclarePairedDelimiter\abs{\lvert}{\rvert}   
\DeclarePairedDelimiter\norm{\lVert}{\rVert}  
\DeclarePairedDelimiter\bkt{[}{]}             
\DeclarePairedDelimiter\brc{\{}{\}}           
\DeclarePairedDelimiter\prn{(}{)}             

\providecommand\giv{}

\alias\given\giv 

\DeclarePairedDelimiterXPP\pp[1]{\mathbb{P}}[]{}{
   \renewcommand\giv{\nonscript\:\delimsize\vert\nonscript\:\mathopen{}}
   #1}
\DeclarePairedDelimiterXPP\ppup[2]{\mathbb{P}^{#1}}[]{}{
   \renewcommand\giv{\nonscript\:\delimsize\vert\nonscript\:\mathopen{}}
   #2}
\DeclarePairedDelimiterXPP\ppdown[2]{\mathbb{P}_{#1}}[]{}{
   \renewcommand\giv{\nonscript\:\delimsize\vert\nonscript\:\mathopen{}}
   #2}
   
\DeclarePairedDelimiterXPP\ee[1]{\mathbb{E}}[]{}{
   \renewcommand\giv{\nonscript\:\delimsize\vert\nonscript\:\mathopen{}}
   #1}

\DeclarePairedDelimiterXPP\eeup[2]{\mathbb{E}^{#1}}[]{}{
   \renewcommand\giv{\nonscript\:\delimsize\vert\nonscript\:\mathopen{}}
   #2}

\DeclarePairedDelimiterXPP\eedown[2]{\mathbb{E}_{#1}}[]{}{
   \renewcommand\giv{\nonscript\:\delimsize\vert\nonscript\:\mathopen{}}
   #2}

\DeclarePairedDelimiterXPP\eeud[3]{\mathbb{E}^{#1}_{#2}}[]{}{
   \renewcommand\giv{\nonscript\:\delimsize\vert\nonscript\:\mathopen{}}
   #3}

\let\preexp\exp
\let\exp\relax

\DeclarePairedDelimiterXPP\exp[1]{\preexp}(){}{#1}

\newcommand{\Exp}[1]{e^{#1}}

\newcommandx{\prf}[2][2=t\in\bkt{0,T}]{ \set{#1}_{#2}}
\newcommandx{\seq}[2][2=n\in\N]{\set{#1}_{#2}}

\newcommandx{\fml}[2]{\set{#1}_{#2}}
\newcommandx{\tol}[1]{\stackrel{#1}{\longrightarrow}}
\newcommandx{\eqd}[1][1=d]{\stackrel{(#1)}{=}}
\newcommandx{\neqd}[1][1=d]{\stackrel{(#1)}{\neq}}

\makeatletter
\let\oldabs\abs \def\abs{\@ifstar{\oldabs}{\oldabs*}}
\let\oldab\ab \def\ab{\@ifstar{\oldab}{\oldab*}}
\let\oldnorm\norm \def\norm{\@ifstar{\oldnorm}{\oldnorm*}}
\let\oldbkt\bkt \def\bkt{\@ifstar{\oldbkt}{\oldbkt*}}
\let\oldbrc\brc \def\brc{\@ifstar{\oldbrc}{\oldbrc*}}
\let\oldprn\prn \def\prn{\@ifstar{\oldprn}{\oldprn*}}
\let\oldpp\pp \def\pp{\@ifstar{\oldpp}{\oldpp*}}
\let\oldppup\ppup \def\ppup{\@ifstar{\oldppup}{\oldppup*}}
\let\oldppdown\ppdown \def\ppdown{\@ifstar{\oldppdown}{\oldppdown*}}
\let\oldee\ee \def\ee{\@ifstar{\oldee}{\oldee*}}
\let\oldeeup\eeup \def\eeup{\@ifstar{\oldeeup}{\oldeeup*}}
\let\oldeedown\eedown \def\eedown{\@ifstar{\oldeedown}{\oldeedown*}}
\let\oldeeud\eeud \def\eeud{\@ifstar{\oldeeud}{\oldeeud*}}

\let\oldexp\exp \def\exp{\@ifstar{\oldexp}{\oldexp*}}
\makeatother

\newcommand{\Bee}[1]{ \ee*[\Big]{#1} }

\newcommand{\Bprn}[1]{ \prn*[\Big]{#1} }
\newcommand{\bprn}[1]{ \prn*[\big]{#1} }

\newcommand{\set}[1]{ \brc{#1} }

\alias{\R}{{\mathbb R}}
\alias{\C}{{\mathbb C}}
\alias{\Z}{{\mathbb Z}}
\alias{\N}{{\mathbb N}}
\alias{\Nz}{{\mathbb N}_0}

\DeclareMathOperator*\esssup{esssup}

\newcommand{\oo}[1]{\frac{1}{#1}}
 
\newcommand{\too}[1]{\tfrac{1}{#1}}
\newcommand{\tot}{\too{2}} 

\renewcommand{\implies}{\rightarrow}
\newcommand{\Implies}{\Rightarrow}

\newcommand{\tint}{\textstyle\int}

\newcommand{\eps}{\varepsilon}
\newcommand{\ld}{\lambda}

\newcommand{\sE}{\mathcal{E}} 
\newcommand{\sF}{\mathcal{F}}

 \newcommand{\bL}{\mathbb{L}}

\newcommand{\sP}{\mathcal{P}}

\newcommand{\sS}{\mathcal{S}}

 \newcommand{\hS}{\hat{S}}

 \newcommand{\hY}{\hat{Y}}
 \newcommand{\hZ}{\hat{Z}}

\newcommand{\lone}{\bL^1}
\newcommand{\ltwo}{\bL^2}

\newcommandx{\upp}[2]{{#1}^{(#2)}}

\newcommand{\eas}{\text{ a.s.}}

\newcommand{\efor}{\text{ for }}
\newcommand{\eforall}{\text{ for all }}

\newcommand{\eand}{\text{ and }}

\newcommand{\ewhere}{\text{ where }}

\newcommand{\itos}{It\^ o's}

\newcommand{\hrho}{\hat{\rho}}

\newcommand{\tv}{\tilde{v}}
\newcommand{\hxi}{\hat{\xi}}
\DeclareMathOperator{\bmo}{bmo}
\newcommand{\fxi}{\bm{\xi}}
\newcommand{\fP}{\bm{P}}
\newcommand{\oV}{\overline{V}}
\begin{document}

\title{Kyle's Model with Stochastic Liquidity}

\date{\today}

\author{Ibrahim Ekren}
\address{Florida State University, Department of
    Mathematics, 1017 Academic Way, Tallahassee, FL 32306}
\email{iekren@fsu.edu}

\author{Brad Mostowski}
\address{Florida State University, Department of
    Mathematics, 1017 Academic Way, Tallahassee, FL 32306}
\email{bm18g@my.fsu.edu}

\author{Gordan \v{Z}itkovi\'{c}}
\address{The University of Texas at Austin,
    Department of Mathematics, 2515 Speedway, Austin, TX 78712}
\email{gordanz@math.utexas.edu}

\thanks{During the preparation of this work the first named author was
supported by the National Science Foundation under Grant No.~DMS-2007826
the third named author by the  National Science Foundation under Grant
No.~DMS-1815017. Any opinions, findings and conclusions or
recommendations expressed in this material are those of the author(s)
and do not necessarily reflect the views of the National Science
Foundation (NSF).
AMS classification (2020): Primary 60H30, 60J60; secondary 91B44
}

\begin{abstract}
We construct an equilibrium for the continuous time Kyle's model with
stochastic liquidity, a general distribution of the fundamental price,
and correlated stock and volatility dynamics. For distributions with
positive support, our equilibrium allows us to study the impact of the
stochastic volatility of noise trading on the volatility of the asset.
In particular, when the fundamental price is  log-normally distributed,
informed trading forces the log-return up to maturity to be Gaussian for
any choice of noise-trading volatility even though  the price process
itself comes with stochastic volatility. Surprisingly, we find that in
equilibrium both Kyle's Lambda and its inverse (the market depth) are
submartingales.
\end{abstract}

\maketitle

\section{Introduction}
Kyle's model, introduced in \cite{kyle}, is one of the most influential
models in the market microstructure literature. The equilibrium
constructed in \cite{kyle} shows how the information about an asset is
incorporated in its price and how the liquidity in the market and the
volatility of the asset price are impacted by noise trading. In its
original formulation, at the initial time, an informed trader learns the
fundamental value $\tilde v$ of an asset, where $\tilde v$ is assumed to
have a normal prior distribution. She then trades against a market maker
in order to optimize her expected profit: The objective of the market
maker, on the other hand, is to filter the fundamental price $\tilde v$
by observing the totality of the demand from the informed trader and one
or more noise traders. To achieve that, she chooses a mechanism which
continuously transforms the observed demand into a price quote. Such a
mechanism, when it satisfies an additional assumption related to market
efficiency, is called an equilibrium if neither the insider nor the
market maker have an incentive to deviate from their pre-announced
strategies.

The equilibrium constructed in \cite{kyle} is linear and the informed
trader's trading rate is proportional to the current price mismatch (the
difference between the quoted and the fundamental price) of asset, and
inversely proportional to the time to maturity. In equilibrium the
increment $dP_t$ of the price is given by $$dP_t=\lambda\, dY_t,$$ where
$dY_t$ is the increment of the total demand received by the market maker
(from the informed trader and the noise traders) and the constant
$\lambda$ is the so-called Kyle's Lambda which is the sensitivity of the
price to the total demand. This constant is proportional the the
standard deviation of the fundamental price and inversely proportional
to the standard deviation of noise trading. Thus, Kyle's model is a
mathematical expression of the idea that the market liquidity is
inversely proportional to the average flow of new information and
proportional to the volume of liquidity-motivated transactions (see
\cite{bagehot1971only} written by Jack Treynor under the pseudonym
Walter Bagehot).

\subsection{Literature review}
An impressive number of extensions of Kyle's model have been considered
in the literature. In discrete time, \cite{subrahmanyam1991risk} allows
for risk aversion of the informed trader, while
\cite{caballe1994imperfect,garcia2020multivariate} work with multiple
assets. In continuous time, \cite{back1992} removes the normality
assumption of the fundamental price and proves the existence of an
equilibrium using a PDE based approach. Kyle's model with dynamic
information is studied in \cite{bp} and \cite{ccd}. We also mention
\cite{back1993,back2018identifying,back2020optimal,biagini2012insider,
bose2020kyle,bose2021kyle,cd1,ccetin2018financial,cho, CHOI201922,
corcuera2020path,cn2,aa2012,barger2021insider,lasserre2004asymmetric}
and \cite{ying2020pre}, among many others.

Recently, \cite{cdf} and \cite{collin2021informed} proposed an extension
of Kyle's model which allows for stochastic volatility in the definition
of the  noise traders' cumulative demand process $Z_t$. In the seminal
work of \cite{kyle}, the instantaneous demand $dZ_t$ of the noise trades
has a deterministic variance $\sigma^2 dt$, where $\sigma$ is either a
constant or  a deterministic function of time. In \cite{cdf} and
\cite{collin2021informed}, however, this variance is given by
$\sigma_t^2 dt$ for some stochastic process $\sigma_t$. Thus, the total
demand $Z_T=\int_0^T dZ_t$ is no longer necessarily Gaussian (as in the
classical case) and it is not clear how the PDE based approach of
\cite{back1992}, or the optimal transport methodology of
\cite{back2020optimal}, needs to be modified in order to find an
equilibrium.

Assuming that the fundamental price $\tilde v$ is normally
distributed an equilibrium is constructed in \cite{cdf}. Relying on
normality, these authors conjecture that the trading rate and the
expected wealth of the informed trader are a linear and a quadratic
function, respectively, of the price mismatch. A crucial step in their
existence proof of the equilibrium is the construction a martingale
whose inverse $\lambda_t$ has the property that the relation $$
dP_t=\lambda_t\, dY_t$$ implies that the conditional variance of $\tilde
v$ decreases at the rate $\lambda^2_t\sigma^2_t$ (as in equation (8) on
p.~1447 of \cite{cdf}). This is accomplished by introducing a
decomposition of $\lambda_t$ which reduces the problem to a Backward
Stochastic Differential Equation (BSDE).

\subsection{Our contributions} Even though it enhances tractability, the
assumption that the fundamental price is Gaussian in
\cite{cdf,collin2021informed} permits the equilibrium price to be
negative with positive probability. One of the goals of the present
paper is to relax this assumption and allow a general distribution for
the fundamental price. The same relaxation in \cite{back1992} renders
the pricing rule non-linear, so we cannot expect the linear-quadratic
structure of \cite{cdf} to apply to our framework either. In fact, in
many aspects the relation between our work and \cite{cdf} is similar to
the relation between \cite{back1992} and \cite{kyle}.

From the control-theoretic perspective, the main contribution of this
paper is the identification of a relevant state process $\xi_t$ that
allows us to compute the price of the asset and the strategy of the
informed trader. It turns out that in equilibrium this state process is
given by $\xi_t=\int_0^t \lambda_s dY_s$, the pricing rule is a function
of $\xi_t$, and the trading rate of the informed trader is proportional
to the deviation of $\xi_t$ from its final value.

Our proof of  existence of the equilibrium relies on the crucial
observation that, in equilibrium, and conditionally on the information
available to the market maker at time $t$, the random variable
$\xi_T-\xi_t=\int_t^T\lambda_s dY_s$ is centered Gaussian with variance
$\int_t^T \lambda^2_s \sigma^2_s ds$ (a quantity which turns out to be
known to the market maker at time $t$). That allows us to use the PDE
based construction of \cite{back1992} and Gaussian filtering to find an
equilibrium pricing rule. The equilibrium price is then of the form $P_t
=H^*(t,\xi_t)$, where $\xi_t$ is the state process mentioned above and
the pricing rule $\xi\mapsto H^*(t,\xi)$ is a random field adapted to
the filtration of $\sigma_t$. It has the property that its section at
maturity, i.e.~the map $\xi\mapsto H^*(T,\xi)$, pushes the Gaussian
distribution of $\xi_T$ to the distribution of $\tilde v$. Moreover, the
random field $H(t,\xi)$ (and its path-dependence in $\sigma_t$) admits a
further simplification as $H(t,\xi)=R_\xi\left(\Sigma_t,\xi\right)$
where $R$ is a deterministic function which solves the heat equation and
$\Sigma_t= \int_t^T \lambda^2_s \sigma^2_s ds$ is the remaining
uncertainty in the final value of $\xi_T$.

The trading rate of the informed trader is proportional to $\tilde
v-\xi_t$ where the constant of proportionality is adapted to the
filtration of the market maker and matches the parallel term in
\cite{cdf}. From the point of view of the market maker the process
$\xi_t$ is a martingale, while for the informed trader it is a bridge
process of type constructed in \cite{cdf}. In fact our process $\xi_t$
has many the features of the price process of \cite{cdf}.

In addition to the relaxation of the Gaussian property of $\tilde v$,
our construction extends the results in \cite{cdf} in several other
ways. First of all, we prove that the strategy of the informed trader is
not only optimal among all absolutely continuous strategies but also
among all strategies with jumps or diffusive components. The optimality
of an absolutely continuous strategy is related to the positivity of the
price impact as mentioned in
\cite{corcuera2020path,bose2020kyle,bose2021kyle},  which holds in our
framework, too.

Next, we allow the cumulative demand process $Z_t$ of the noise trades
and its stochastic volatility $\sigma_t$ to be driven by correlated
Brownian motions. As mentioned in \cite{cdf} in the Gaussian case, the
fact that $\sigma_t$ is observable by both agents implies that $d\xi_t$
is not driven by $dY_t$ but by the innovation process for the filtering
problem of the market maker. This innovation process is  orthogonal to
the increments of $\sigma_t$ even if $\sigma_t$ and $Z_t$ are driven by
correlated Brownian motions. Therefore, quite surprisingly, $\sigma_t$
and $\xi_t$ (and therefore $P_t$) are driven by independent Brownian
motions in equilibrium.

Another interesting finding concerns Kyle's Lambda, i.e.~the sensitivity
of the price to the total demand $Y_t$ (or its informative part $\hY$ when
$\sigma_t$ and $Z_t$ are driven by correlated Brownian motions).  In
equilibrium it is given by
$\frac{R_{\xi\xi}(\Sigma_t, \xi_t)}{\frac{1}{\lambda_t}}$ which is a ratio of
two positive orthogonal martingales for the filtration of the market
maker. Thus, Kyle's Lambda is a submartingale.  Trivially, but also
surprisingly, the market depth which is the inverse of Kyle's Lambda is
also a submartingale as the ratio of two orthogonal positive
martingales. With Gaussian distributions as in \cite{cdf}, the function
$R_{\xi\xi}$ is constant and Kyle's Lambda is a submartingale, but the
market depth - being equal to $\frac{1}{\lambda_t}$ up to a
multiplicative constant - is a martingale.

Under the assumption that the risk neutral and physical probabilities
agree (as they do in the context of Kyle's model with a risk-neutral
market maker), we can use the full set of call option prices to gain
information about the distribution of $\tilde v$. Indeed, given a choice
of dynamics for $\sigma_t$, we can use our model to predict the dynamics
of the implied volatility curve for a given maturity as a function of
$\xi_t$ and $\Sigma_t$. With observed call option prices used as input,
this leads to an inverse problem for the distribution of $\tilde v$. For
example, if the distribution of $\tilde v$ is log-normal, i.e.~ with a
flat IV (implied volatility) curve, then its IV curve remains flat on
$[0,T]$. However, the level of this flat curve moves stochastically
depending on the value $\sigma_t,\xi_t$ and $\Sigma_t$. For general
distributions of $\tilde v$, the shape of the IV curve might change
depending on time, $\xi_t,\sigma_t$ and $\Sigma_t$ in a nonlinear way up
to the computation of $R_\xi$ by solving a heat equation and inverting
the Black-Scholes formula as a function of the volatility.
Qualitatively, we observe that the shape of the IV curve is mainly
influenced by the distribution of $\tilde v$ whereas its level mainly
depends on the dynamics of $\sigma_t$.

As mentioned in \cite{cdf}, the market maker anticipates more informed
trading when there is more noise trading. Thus, the rate of injection of
the information into the asset price is stochastic. This effect imposes
distributional constraints on the dynamics of the prices process. For
example, if $\tilde v$ has a lognormal distribution $\xi_t$ can be
identified as the log-return of the asset price up to a multiplicative
constant. Thus, the fact that $\int_t^T \lambda^2_s \sigma^2_s ds$ is
measurable with respect to the information of the market maker at time
$t$ means that independently of the stochasticity of future noise
trading, the presence of an informed trader renders the log-return of
the asset from $t$ to $T$  Gaussian. Note also that the distribution of
$\tilde v$ does not impact $\lambda_t$ so that the (conditional)
Gaussianity of $\xi_T$ still holds for general distributions. If the
fundamental price is Gaussian, $\xi_t$ is the price process up to a
multiplicative constant and the adaptedness of $\int_t^T \lambda^2_s
\sigma^2_s ds$ to the information of the market maker imposes a centered
Gaussian conditional distribution onto the the price increment
$P_T-P_t$. For general distributions, $\xi_t$ can not be interpreted as
either the return of the asset or its price and the dependence between
price process and $\xi$ is nonlinear in general.

\subsection{Organization of the paper.} The rest of the paper is
organized as follows. In Section \ref{s.notations}, we first state the problem
and define the concept of equilibrium. Then we introduce its most
important building blocks and state our main existence result.
Section \ref{s.examples} provides examples and Section \ref{s.proofs}
contains the proofs of the main theorem and other results.

\section{Problem setup and the main result}\label{s.notations}

\subsection{The probabilistic setup}
Let $T>0$ and let $(\Omega, \cF,(\cF_t)_{t\in[0,T]}, \P)$ be a filtered
probability space satisfying the usual conditions of right continuity
and completeness.  We suppose that $(\cF_t)$ is the right-continuous
filtration given by $\sF_t = \sF^{W,B}_t \vee \sigma(\tv)$, where
$\sF^{W,B}$ is the usual augmentation of the filtration generated by $W$ and $B$. More generally, 
for any process $S$, we denote by
$\sF^{S}$ the the augmentation of the filtration generated by $S$ (in
this context, we interpret $\tv$ as a constant process so that, e.g., $\sF^{\tv,W,B}_t = \sF_t$). 

For an $\cF^W$-martingale $M$, we define its BMO (bounded mean
oscillation) norm by
$$ ||M||^2_{BMO}=\sup_{\tau} ||\E[|
M_T-M_\tau|^2|\cF^W_{\tau}]||_\infty$$ where the supremum is taken over
all $\cF^W$-stopping times $\tau$, and $||\cdot||_\infty$ denotes the
essential supremum of a random variable (see
\cite{kazamaki2006continuous}). We call $M$ a BMO-martingale if
$||M||_{BMO}<\infty$. For any $\sF^W$-adapted and $W$-integrable process
$\a$, we write $\a \in \bmo$ if $\int \a_s dW_s$ is a BMO-martingale.
For a continuous process $M$ and $\gamma\in (0,1)$, we denote by
$|M|_\gamma:=\sup_{0\leq s<t\leq T}\frac{|M_s-M_t|}{|t-s|^\gamma}$, its
pathwise $\gamma$-Holder semi-norm. 

Let $\cS^{\infty}$ denote the set of continuous, $\cF^W$-adapted
and uniformly bounded processes. The set
$\cS^{+}_0$ consists of all
continuous $\cF^W$-adapted processes $G$, strictly positive on $[0,T)$,
with $G_T=0$, and $\cP^2$ denotes the set of all
$\cF^W$-progressively measurable processes $z$ with $\int_0^T z_u^2\,
du<\infty$, a.s.

\subsection{The model}
As in \cite{cdf}, we consider an interaction among an informed
trader, a market maker and noise traders 
during the time period $[0,T]$:
\begin{itemize}[label=-, leftmargin=3ex]
\item
 At time $t=0$, the informed trader (insider) learns the value of $\tilde
v$, a random variable that represents the fundamental (or liquidation)
value of an asset at maturity $t=T$.  He trades in the market using a
strategy $X$, where $X$ denotes the total cumulative demand. We allow
$X$ to depend $\tv$, as well as both $W$ and $B$ in an adapted way,
i.e., the insider's filtration is $\sF$. 
\item Noise traders place their trades at random without any regards
to the actions of the other participants. Their trading intensity is not
constant, but given by a stochastic process $\sigma$, so that
the cumulative order process $(Z_t)$ of the noise traders is 
given by
\begin{align}\label{eq:defZ}
    Z_t=\int_0^t \sigma_s \Bprn{\hrho\, dB_s+\rho\, dW_s}, t\in [0,T],
    \ewhere \rho \in (-1,1) \eand \hrho=\sqrt{1-\rho^2}.
\end{align}
\item The market maker has no access to the value of $\tilde v$ or the
demand $X$ of the insider. On the other hand, she knows the distribution
$\nu$ of $\tv$ and observes the total order flow $Y = X + Z$, as well as
the value of $W$; in other words, her filtration is given by $\sF^m =
\sF^{Y,W}$. She precommits to a pricing functional $\fP$ which 
transforms the entire
observed path of $Y$ and $W$, as in \cite{cdf}, to a price process $P$. 
\item The equilibrium is achieved when the insider has no incentive to
alter his trading strategy $X$, given the pricing functional $\fP$, and the
market maker's price $P = \fP(X+Z)$ is rational 
(i.e., the $\ltwo$-optimal estimate of $\tv$ based on her information) 
given the insider's strategy $X$.
\end{itemize}
We proceed by giving rigorous definitions for the concepts introduced
informally above:
\begin{definition}\label{pricing-rule}
	A \define{pricing rule} is a map $\fP$ that assigns to
each $\sF$-semimartingale $S$ an $\cF^{W,S}$-semimartingale $\fP(S)$ in a
nonanticipative manner, i.e., for each $t\in [0,T]$ we have
\begin{align*}
  S_s = S'_s \eforall s\leq t,\eas \ \Implies 
  \ \fP(S)_s = \fP(S')_s \eforall s\leq t, \eas.
\end{align*}
\end{definition}

\begin{remark}
Our equilibrium price functional $\fP^*$ will be built in two steps.
First, a state process $\xi^*$ will be constructed by applying a 
non-anticipative functional $\fxi^*$ of the paths of $Y$ and $W$. 
Then, a random field $H^*$ adapted
to $\sF^W$ will be applied to it: Compared to \cite{back1992}, we
interpret $\xi^*$ as a path dependent generalization of the total demand
process $Y$, while $H^*$ adds $W$-dependence to Back's $H$. 
In this regard, $\xi^*$ is a novel state variable allowing us to
state the equilibrium as a one dimensional Markov control problem (of
$\xi$) from the perspective of the informed trader and the pricing rule
as a functional of $(\xi_t)$. We refer to
\cite{cho,ccd,bose2020kyle,bose2021kyle} for the introduction of
auxiliary state processes in Kyle's model. 
\end{remark}

Once the price functional $\fP$ is given, the insider's goal is to
maximize the expected gains from investing in the market. To rule out
doubling strategies and other pathologies, we impose an admissibility
constraint in the standard way. We recall from \cite{back1992}, that the
total profit/loss from trading accumulated by the informed trader who
uses the strategy $X$ against the price process $P$ is given by $(\tv -
P_T) X_T + \int_0^T X_{t-}\, dP_t$. The same author uses integration by
parts to cast this expression into an equivalent, but more convenient form $\int_0^T (\tv
- P_s)\, dX_s - [X,P]_T$, which we use the definition of admissibility
below:
\begin{definition}
	Any $(\cF_t)$ semimartingale $X$ with $X_0=0$ is called a
	\define{trading strategy}. Given a semimartingale $P$, and a trading
	strategy $X$, the random variable  
	\begin{align} \label{Pi}
        \Pi(X,P)_T = (\tv - P_T) X_T + \int_0^T X_{t-}\, dP_t 
        = \int_0^T (\tv - P_t)\, dX_t - [X,P]_T 
	\end{align}
is called the \define{realized wealth} of the strategy $X$, with
respect to $P$.

Given a semimartingale $P$, a trading strategy $X$ is said to be
\define{$P$-admissible} if the process $\Pi(X,P)_t =  \int_0^t (\tv -
P_s)\, dX_s - [X,P]_t$ is uniformly bounded from below by an integrable
random variable.

Given a pricing functional $\fP$, 
a trading strategy $X$ is said to be \define{$\fP$-admissible} if it is
$\fP(X+Z)$-admissible.
\end{definition}

\begin{remark}
Note that unlike \cite{cdf}, we allow the informed trader to use both
diffusive and jump strategies. However, we prove in the sequel that
these strategies are not profitable for the informed trader and in
equilibrium it is optimal for the informed trader to use an absolutely
continuous strategy. As noted in
\cite{back2020optimal,corcuera2020path}, this point is inherited from
the fact that the final pricing rule of the market maker is an
increasing function of the underlying state variable.
\end{remark}

Finally, we introduce the standard notion of  equilibrium:
\begin{definition}
	A pair $(\fP^*, X^*)$ consisting of a pricing rule $\fP^*$ and a
	trading strategy $X^*$,
	is an \define{equilibrium} if
	\begin{itemize}
    \item[(i)] $X^*$ is $\fP^*$-admissible  and
        \begin{align*}
            \Bee{\Pi\Bprn{X;\fP^*(X+Z)_T}} \leq \Bee{ \Pi\Bprn{X^*;\fP^*(X^*+Z)_T}},
        \end{align*}
        whenever $X$ is an $\fP^*$-admissible trading strategy.
    \item[(ii)] $\fP^*$ is rational i.e., 
	\begin{align}\label{rational}
	\fP^*(X^*+Z)_t = \ee{ \tv \giv \cF^{W, X^*+Z}_t}, \eas, 
    \eforall t\in [0,T]. 
	\end{align}
\end{itemize}
\end{definition}
\begin{remark}
Notationally, we distinguish between functionals (bold, like $\fP$) and
processes (light, like $P$). Similarly, starred quantities (like $X^*$)
will refer to the (candidate) equilibrium, while their non-starred
versions (like $X$) denote their generic analogues. The two notations
are often used together (as in $\fxi^*$).
\end{remark}

\subsection{Regularity assumptions.}
Before we state our main result, we discuss the regularity assumptions
imposed on its inputs. Examples of processes which satisfy 
part \eqref{it:sigma} will be provided in Section \ref{s.examples}
below.
\begin{assumption} \ \vspace{-0.5em}
    \label{asm:sigmanu}
\begin{enumerate}
   \item \label{it:rho} $|\rho|<1$,
   \item \label{it:tv} $\tv \in \bL^2$ and its distribution  
   $\nu$ is absolutely continuous.
   \item \label{it:sigma} $\sigma$ admits a decomposition of the form
   \begin{align}\label{sLJ}
     \sigma = LJ,
   \end{align}
where $L$ is a stochastic exponential of a BMO martingale adapted to $\sF^W$,
and $J$ is an $\sF^W$-adapted, bounded and
bounded-away-from-$0$ continuous process with the property that
\begin{align}\label{eq:boundholder}
    \E[e^{r |J|_\gamma}]<\infty\mbox{ for all } r>0.
\end{align}
for some H\" older exponent $\gamma > 0$.
\end{enumerate}
\end{assumption}

\begin{remark}\label{rmk:cdf}
It is readily seen that the requirements of Assumption \ref{asm:sigmanu},
\eqref{it:sigma} are satisfied for a volatility process with the decomposition
\begin{align*}
  d\sigma_t = \sigma_t \prn{ b_t\, dt + \psi_t\, dW_t}, \ \sigma_0>0
\end{align*}
where $b$ and $\psi$ are $\sF^W$-adapted, $b$ is bounded and $\psi \in
\bmo$ (or, more restrictively, bounded as well). 
In general, this condition is more stringent
than the assumptions imposed on $\sigma$ in \cite{cdf}. This is due the
fact that we aim to fix few minor mistakes in \cite{cdf}. Indeed, Lemma
8., p.~1471 in \cite{cdf} states the martingality of two processes for
any admissible strategy of the informed trader. This claim is proven by
using Lemma 4., p.~1468 in \cite{cdf}. Unfortunately, this Lemma only
applies to the candidate optimal strategy of the informed trader and not
to all admissible strategies. {Additionally due to $\Sigma_T=0$, the martingality of the process in \cite[Equation (66)]{cdf} requires additional arguments. Fixing these minor mistakes for a reasonable class
of admissible strategies turns out to be a somewhat challenging problem
that requires the more stringent Assumption \ref{asm:sigmanu}.}
\end{remark}

\subsection{Building blocks of the equilibrium}
\label{su:building-blocks}
The main goal of this paper is to show that, under Assumption
\ref{asm:sigmanu} an equilibrium exists,  and to describe its structure.
We outline its construction here, with all proofs left for section
\ref{s.proofs}. 

\subsubsection{The function $h$.}
Let $h$ be the unique nondecreasing function that pushes the standard normal distribution
forward to the distribution $\nu$ of $\tv$. More precisely, let
$F_{\nu}$ be the cdf (cumulative distribution function) of $\nu$ with $F^{-1}_{\nu}$ it generalized inverse, let
$\Phi$ be the cdf of the standard normal and let 
\[ h(x) = F^{-1}_{\nu}(\Phi(x)) \efor x\in \R.\]
It is clear that $h$ is nondecreasing and right-continuous, and unique
(in the class of nondecreasing functions) with the property that $h_\sharp N(0,1)  = \nu$, 
where $h_\sharp$ denotes the push-forward and 
$N(\mu,\sigma^2)$ denotes the normal distribution with mean $\mu$ and
variance $\sigma^2$. Moreover, since $\nu$ is absolutely continuous,
its inverse $h^{-1}$ is well-defined, and the random variable
$h^{-1}(\tv)$, 
has the standard normal distribution and the
property that $h(h^{-1}(\tv)) = \tv$, a.s.

\subsubsection{The function $R$.}
Let $p(u,\cdot)$ denote the
probability density function (pdf) of $N(0,u)$ for $u\in (0,1]$; 
recall that $p$ is the fundamental solution of the heat equation on
$\R$.
Lemma \ref{nine-part}, \eqref{it:h-int}, guarantees that the function
$R:[0,1] \times \R \to \R$, given by 
\begin{align} \label{R-FK}
    R(t,\xi) = \int \prn{\tint_0^{\xi+\zeta} h(x)\, dx} p(t,\zeta)\, d\zeta , 
\end{align}
is well defined. 
Moreover, it belongs to the class $C^{1,2}((0,1]\times \R) \cap C([0,1]\times
\R)$ and solves the following initial-value problem (see, e.g.,
\cite[Section 4.3, p.~254]{KarShr91} for details)
\begin{equation}\label{eq:heat}
    \left\{
    \begin{aligned}
        R_t &= \tot R_{\xi\xi},\ (t,\xi) \in (0,1)\times \R, \\
        R(0,\xi) &= \tint_0^{\xi} h(x)\, dx,\ \xi \in \R.
    \end{aligned}
    \right.
\end{equation}
The dominated convergence theorem implies that the equation \eqref{R-FK}
can be differentiated under the integral sign and that the derivative
$R_{\xi}$ solves an initial problem for the heat equation, too, 
but with the terminal condition $R_{\xi}(0,\xi) = h(\xi)$. Note that the
monotonicity of $h$ implies that $R(t,\cdot)$ is convex and
$R_{\xi}(t,\cdot)$ nondecreasing.

\subsubsection{Processes $G$, $\Sigma$ and $\ld$.}
The core of the argument needed for the construction of processes 
$\Sigma$ and $\ld$ is given in the following proposition, whose proof is
postponed until section \ref{s.proofs}.
\begin{proposition}\label{G}
 Under Assumption \ref{asm:sigmanu}, 
the backward stochastic differential equation (BSDE)
\begin{align}\label{bsdeG}
    G_t & = \int_t^T \Bprn{\hrho^2\sigma_s^2 -\frac{U_s^2}{4G_s}}
    \, ds -\int_t^T U_s\,dW_s
\end{align}
admits a solution $(G,U)$, unique in the class $\sS^+_0 \times \sP^2$.
This solution has the following properties:
\begin{align}\label{G-prop}
  \frac{U}{G}\in\bmo \eand \int_0^T \frac{\sigma^2_s}{G_s}\, ds = +\infty, \eas
\end{align}
\end{proposition}
Using the process $G$ of Proposition \ref{G}
above, we define two more $\sF^W$-adapted processes
\begin{align}\label{Sigma-ld}
  \Sigma_t = \exp{ - \int_0^t \frac{\hrho^2 \sigma_s^2}{G_s}\, ds} 
  \efor t<T \eand 
  \ld_t = \frac{\sqrt{\Sigma_t}}{\sqrt{G_t}},
\end{align}
noting that, by the second statement in \eqref{G-prop}, 
the extension $\Sigma_T=0$ makes $\Sigma$
continuous. By \itos{} formula we have
\begin{align}\label{ld-dyn}
  d \prn{\oo{\ld_t}} = \oo{\ld_t} \frac{U}{2G}\, dW_t,
\end{align}
and, by the first statement in \eqref{G-prop}, $\oo{\ld}$ is an $\sF^W$-martingale.

\subsubsection{The functional $\fxi^*$ and the pricing rule $\fP^*$}
With $G$, $U$ and $\lambda$ at our disposal, we are ready to define 
the candidate pricing rule.  First, we define the functional $\fxi^*$
which acts on an $\sF$-se\-mi\-mar\-tin\-gale $S$ as follows:
\begin{align}\label{def:state}
    \fxi^*(S)_t=\int_0^t\lambda_s \prn{d\hS_s-\frac{U_{s}}{2 G_{s}} 
    d[ \hS,W]_s}\,  \ewhere \hS_t = S_t - \int_0^t \rho \sigma_s\, dW_s.
\end{align}
Since $\ld$ is continuous and $U/G\in \bmo \subseteq \sP^2$ the integral in
\eqref{def:state} exists a.s., and defines an 
$\sF^{W,S}$-semimartingale for any semimartingale $Y$, in a
nonanticipating way. 

\noindent With $\fxi^*$ at hand, we define the candidate pricing
rule $\fP^*(S)$ as a composition:
\begin{align}\label{def-P} \fP^*(S)_t = H^*\bprn{t, \fxi^*(S)_t} 
    \ewhere H^*(t,\xi) = R_{\xi}(\Sigma_t,
\xi) \efor (t,\xi) \in [0,T]\times \R.
\end{align}
The adaptivity properties of $H^*$ and $\fxi^*$ imply that $\fP^*$ is
indeed a pricing rule in the sense of Definition \ref{pricing-rule}
above. 

\subsubsection{Processes $\xi^*$ and $X^*$}
\label{xi-sub}
{We prove in subsection \ref{su:proof} below that
there exists 
a  unique process $\xi^*_t$ which is continuous on $[0,T]$ and satisfies
\begin{align}\label{def:state2}
    \xi^*_t =\int_0^t   \frac{\lambda^2_s \hrho^2 \sigma^2_s }{\Sigma_s }
    \prn{h^{-1}(\tilde v)-\xi^*_s}+\hat \rho\int_0^t \lambda_s\sigma_s dB_s \efor t\in [0,T).
\end{align}}
The process $\xi^*$ is, in turn, used to 
define the candidate equilibrium trading strategy of the informed trader as
\begin{align}\label{X-xi-star}
    X^*_t:=\int_0^t \frac{\lambda_s \hrho^2 \sigma^2_s }{\Sigma_s }
    \prn{h^{-1}(\tilde v)-\xi^*_s}\, ds,\ t\in [0,T].
\end{align}
so that by definition of $X^*$ and $\xi^*$ and denoting
\begin{align*}
   Y^*_t = X^*_t + Z_t \eand \hY^*_t = X^*_t + \int_0^t \hrho \sigma_s\, dB_s = 
   Y^*_t - \int_0^t \rho \sigma_s\, dW_s,
\end{align*}
we also have
$$\xi^*_t= \fxi^*(Y^*)_t.$$

\subsection{The main theorem and some properties of the equilibrium}
With all the main building blocks defined and the notation introduced in 
subsection \ref{su:building-blocks} above, we are ready to state our
main result. We remind the reader that
$\sF^{m*}=\sF^{Y^*,W}$ corresponds to the information
available to the market maker, that $\eqd$ denotes the equality 
in distribution and that $h_* \mu$ denotes the push-forward of the
measure $\mu$ by the function $h$.
The convex conjugate $R^c$ of $R$ is defined by
\begin{align}\label{Rc}
 R^c(u,v) = \sup_{\xi\in\R} \Bprn{ \xi v - R(u,\xi)}, \efor  u\in [0,1] 
 \eand v\in \R.
\end{align}
\begin{theorem}\label{thm:main}
    Under Assumption \ref{asm:sigmanu}, the pair
    $(\fP^*, X^*)$ constructed in subsection 
    \ref{su:building-blocks} above,  is an equilibrium.

    Additionally, in that equilibrium, 
\begin{enumerate}
\item $\tv$-conditional (i.e., time $0+$) expected profit/loss of the
informed trader is
$$\frac{R(1,0)+R^c(0,\tv)}{\lambda_0}$$
\item There exists an $\cF^{m*}$-Brownian motion $\hat B$ orthogonal to $W$ so that
\begin{align}
 d\hY^*_t&=\hat \rho \sigma_t\, d\hat B_t,\label{hYs}
\\
\label{xis-hYs}
  d\xi^*_t &= \ld_t\, d\hY^*_t,\eand \\
  \label{Ps-hYs}
  dP^*_t &= \frac{R_{\xi\xi}(\Sigma_t\, \xi^*_t)}{ \oo{\ld_t} }\, d\hY^*_t.
\end{align}
Therefore, the processes $\xi^*$, $Y^*$ and $\hY^*$ are  
$\sF^{m*} $-martingales, with  $\xi^*$ and $\hY^*$ orthogonal  to $W$. 
\item $\xi^*_T=h^{-1}(\tilde v)$ a.s., and, conditionally on $\sF^{m*}_t$, we have $\xi^*_T \eqd N(\xi^*_t, \Sigma_t)$.
Thus, the $\sF^{m*}_t$-conditional distribution of $\tv$ is $h_\sharp
N(\xi^*_t, \Sigma_t)$.
\end{enumerate}
\end{theorem}

\begin{remark}\ 
\begin{enumerate}

\item The $\sF^{m*}$-martingale property of $\xi^*$, the fact that
$\xi^*_T = h^{-1}(\tv)$ and the definition
of the optimal strategy $X^*$ imply that 
\begin{align}\label{inconsp} 
    \ee{ \frac{dX^*_t}{dt} \giv \cF^{m*}_t}=0 \text{ for almost every }
t\in [0,T].\end{align}
This is the "inconspicuous informed trading" property in \cite{cho}; at
each time, the trading intensity of the insider has zero expectation
from the point of view of the market maker.

\item On the filtration $\cF$, $\xi^*$ has the same dynamics as the price
process in \cite{cdf} and therefore it is of new class of bridges
constructed in \cite{cdf}. The final condition of this bridge is
$h^{-1}(\tilde v)$ and the price at maturity is $\tilde v$. Thus, all
information is incorporated to the price at time $T-$. In fact all
qualitative properties of the price process in \cite{cdf} hold for
$\xi^*$ in our context.

\item The trading rate of the informed trader is proportional to
mismatch $(h^{-1}(\tilde v)-\xi^*_t)$ between the value of
$\xi^*_T=h^{-1}(\tilde v)$ which is determined by the private
information of the informed trader and the current value of the state
process $\xi^*_t$. The proportionality term $\hrho^2 \lambda_t
\sigma^2_t \Sigma^{-1}_t$ is $\cF^m_t$ measurable and explodes as $t\to T$
due to the equality $\Sigma_T=0$. This term has the property that
if the market maker solves a filtering problem, the $\cF^m_t$-conditional
distribution of $h^{-1}(\tilde v)$ is $N(\xi^*_t,\Sigma_t)$.

\item It is standard to define Kyle's Lambda as the sensitivity of the
price to the total-demand process $Y^*$.  In our context, the natural
quantity is not $Y^*$ but the adjusted order flow $\hY^*_t = Y_t^*-\int_0^t
\rho \sigma_s dW_s$ which is the innovation process from the perspective
of the market maker. Thus, in our context Kyle's Lambda is given by 
$$\frac{R_{\xi\xi}(\Sigma_t,\xi^*_t)}{\frac{1}{\lambda_t}}$$
which is positive thanks to the convexity of $R$ in $\xi$.

\item Similarly to the computation leading to \eqref{Ps-hYs} above, we
can show that
$$dR_{\xi\xi}(\Sigma_t,\xi^*_t)=
\frac{R_{\xi\xi\xi}(\Sigma_t,\xi^*_t)}{\frac{1}{\lambda_t}}\, d\hY^*_t$$
In equilibrium, $Y^*$ is a
martingale in the filtration $\cF^{m*}$ of the market maker and
Kyle's Lambda becomes the ratio of two $\cF^{m*}$-local
martingales, $R_{\xi\xi}(\Sigma_t,\xi^*_t)$ and $\frac{1}{\lambda_t}$.
As show in Lemma \ref{nine-part}, \eqref{it:orthog} below, 
these two
positive local martingales are orthogonal to each other. Thus, when
$R_{\xi\xi}(\Sigma_t,\xi^*_t)$ is a true martingale, similarly to
\cite{cdf}, Kyle's Lambda is in fact a submartingale and therefore
increasing on average.

In agreement with \cite{cdf}, the submartingality of Kyle's Lambda in
our framework is in contrast with
\cite{back1992,baruch2002insider,bose2020kyle,cho,caldentey2010insider}
where Kyle's Lambda decreases with time and the informed trader suffers
less from adverse selection of her traders close to the maturity.

\item It is also standard to define the market depth as the inverse of
Kyle's Lambda. In \cite{cdf}, due to Gaussianity assumption of $\nu$,
$R_{\xi\xi}$ is a constant. Thus, the market depth process is a
proportional to $\frac{1}{\lambda_t}$ and is a $\cF^{m*}$-martingale. For
general $\nu$, in our context, the market depth is also a submartingale
as the ratio of two orthogonal $\cF^{m*}$ martingales.

\item
The introduction of processes $G$, $\Sigma$ and $\ld$ is an
important contribution of \cite{cdf} in the understanding of Kyle's
models. In particular, we interpret the $\sF^W$-measurable process $\lambda$ as a way of changing the conditional distribution of the
underlying state process. Indeed, although we are unable to describe explicitly the equilibrium distribution of the original state process $Y_T$ conditional to the information of the market maker at time $t$, due to the choice of $\lambda$, $\int_t^T \hrho^2 \lambda_s^2 \sigma_s^2\, ds=\Sigma_t$ is known by the market maker at time $t$. Thus, in equilibrium, $ \xi^*_T$ has a Gaussian distribution with mean $ \xi^*_t$ and variance $\Sigma_t$ conditionally on the information of the market maker. Thus, by integrating $Y$ against $\lambda$ in the definition of the novel state process $\xi^*$ in \eqref{def:state}, we render the novel state process conditionally Gaussian which allows us to use the optimal transport tools of \cite{cel2020}. 

\end{enumerate}
\end{remark}
\section{Examples} \label{s.examples}
We split our examples into two groups. In subsection \ref{su.tv} we treat several
different distributions $\nu$ of the fundamental value $\tv$ and
investigate the shape of the corresponding implied volatility (IV) curve.
Subsection \ref{su.rough} contains a descriptions of a rough 
volatility model based on a stochastic Volterra
equation that satisfies our regularity assumption, Assumption
\ref{asm:sigmanu}, \eqref{it:sigma}. 

\subsection{Distributions of the fundamental value and 
implied-volatility curves}\label{su.tv}
\subsubsection{Normal belief of the market maker}\label{ss.lognormal} If
$\nu$ is Gaussian $N(m_\nu,\sigma_\nu^2)$, then $h$ defined in
Assumption \ref{asm:sigmanu} is
$$h(\xi)=\sigma_\nu \xi+m_\nu \mbox{ and }R_{\xi\xi}(s,\xi)=\sigma_\nu .$$
Thus, $$P_t=\sigma_\nu \xi^*_t+m_\nu$$ and we recover the equilibrium in \cite{cdf}.

In equilibrium, $dP_t=\sigma_\nu \hrho\lambda_t \sigma_t
d\hat B_t$ has a stochastic diffusion term. However, (3) of Theorem \ref{thm:main} leads to the fact that conditionally on $\cF^{m*}_t$,
$$P_T-P_t=\sigma_\nu \hrho\int_t^T\lambda_s \sigma_s d\hat B_s$$
is Gaussian with mean $0$ and variance $\sigma_\nu^2 \Sigma_t$.

\subsubsection{Log normal belief of the market maker}
If $\nu$ is the distribution $m\exp{-\frac{\sigma_v^2}{2}+\sigma_v G}$ for $G$ that is a standard normal distribution. Then,
$$h(x)=m\exp{-\frac{\sigma_v^2}{2}+\sigma_vx}\mbox{ and }R_\xi(t,\xi)=m\exp{\frac{\sigma_v^2}{2}(t-1)+\sigma_vx}.$$
We can compute the price as
\begin{align}\label{eq:lognp}
    P_t=R_\xi(\Sigma_t,\xi^*_t)=m\exp{\frac{\sigma_v^2}{2}(\Sigma_t-1)+\sigma_v\xi^*_t}.
\end{align}

Differentiating \eqref{eq:lognp}, we obtain the dynamics
$$
    \frac{dP_t}{P_t}=\sigma_vd\xi^*_t=\sigma_v\hrho\lambda_t \sigma_t d\hat B_t
$$
where $\hat B$ is defined in \eqref{hYs}.
Thus, $\frac{dP_t}{P_t}$ defines a martingale which as quadratic variation $\sigma_v^2 (1-\Sigma_t)$ and which is orthogonal to $W$. Note that in equilibrium, we obtain a stochastic volatility dynamics for the price where the increments of the price and volatility are orthogonal.

We obtain the log-return $\ln\frac{P_T}{P_t}={-\frac{\sigma_v^2}{2}\Sigma_t+\sigma_v(\xi^*_T-\xi^*_t)}$. Note that $\Sigma_t$ and $\int_t^T\lambda^2_s\sigma^2_sds$
are $\cF^{m*}_t$ measurable. Thus, conditionally on $\cF^{m*}_t$ the log-return is Gaussian with mean $-\frac{\sigma_v^2}{2}\Sigma_t$ and variance $\sigma_v^2(1-\rho^2)\int_t^T\lambda_s^2 \sigma_s^2ds=\sigma_v^2\Sigma_t$ which means that informed trading imposes distributional constraints on the distribution of the log-returns.

Additionally, conditionally on $\cF^{m*}_t$, the distribution of $P_T$ is log normal and the price of a call option with maturity $T$ and strike $K$ can be computed as
\begin{align*}
    \E[(h(\xi^*_T)-K)^+|\cF^{m*}_t] & =\E\left[\left(R_\xi(\Sigma_t,\xi^*_t)\exp{-\frac{\sigma_v^2}{2}\Sigma_t+\sigma_v (\xi^*_T-\xi^*_t)}-K\right)^+|\cF^{m*}_t\right] \\
                                 & =BS\left(t,R_\xi(\Sigma_t,\xi^*_t),T,K,\sqrt{\frac{\Sigma_t}{T-t}}\right)
\end{align*}
where $BS(t,p,T,K,\hat \sigma)$ is the Black Scholes price of the call option where the volatility between $t$ and $T$ is $\hat \sigma$.
Thus, in our model, with a lognormal fundamental price, the IV curve remains flat at each time $t\in[0,T]$ and equal to $\sqrt{\frac{\Sigma_t}{T-t}}$.


\subsubsection{Non-flat IV curve at initial time}
For general distributions of the fundamental prices, it is not possible to solve the heat equation \eqref{eq:heat} explicitly. However, we can still rely on the Feynman-Kac representation to express the equilibrium price as
$$P_t=\E[h(\xi^*_T)|\cF^{m*}_t]=\frac{1}{\sqrt{2\pi}}\int h(\xi^*_t+\sqrt{\Sigma_t}y)e^{-\frac{y^2}{2}}dy$$
where $h$ is not necessarily exponential or linear as in the previous sections.

In fact, the distribution of $\xi^*_T$ conditional on $\cF_t^{m*}$ is $N(\xi^*_t,\Sigma_t)$, we can price any option with payoff $p\mapsto H(p)$ as
\begin{align}\label{optionprice}
\E[H(h(\xi^*_T))|\cF^{m*}_t]=\frac{1}{\sqrt{2\pi}}\int H(h(\xi^*_t+\sqrt{\Sigma_t}y))e^{-\frac{y^2}{2}}dy
\end{align}
and in particular obtain the dynamics of the IV curve at maturity $T$ for any $t,\xi^*_t,\Sigma_t$.

\subsubsection{Gaussian mixtures for returns}
We suppose that the distribution $\nu$ of $\tilde v$ is the mixture of log-normal distributions given by $ \{X_{m_i,\sigma_i};w_i\}_{i=1}^N$, where
\begin{align*}
    X_{m_i,\sigma_i} & = m_i\exp{-\frac{1}{2}\sigma_i^2 + \sigma_iZ}
\end{align*}
with $Z$ being a standard normal random variable and $w_i\geq 0$ are weights satisfying $\sum_{i=1}^N w_i=1$. This is equivalent to assuming that conditional on the choice of an index $i$ with probability $w_i$, the log-price $\log(\tilde v)$ will be given by $ \ln(m_i) - \frac{1}{2}\sigma_i^2 + \sigma_iZ$, i.e. the log-price is a Gaussian mixture. Then, the pdf of $\nu$ is given by
\begin{align*}
    f_\nu(x) = \sum_{i=1}^N \frac{w_i}{\sigma_ix}\phi\Big(\frac{\ln(\frac{x}{m_i}) - \frac{1}{2}\sigma_i^2}{\sigma_i}\Big)
\end{align*}
where $\phi$ is the density function of the standard normal distribution. 
This means the transport map $h(x)$ satisfies
\begin{align*}
    \Phi(x) = \sum_{i=1}^N \frac{w_i}{\sigma_ix}\phi\Big(\frac{1}{\sigma_i}\ln{\Big(\frac{h(x)}{m_i}\Big)} - \frac{1}{2}\sigma_i\Big)
\end{align*}
and we have a simple way of numerically computing $h(x)$.

\begin{figure}%
    \centering
    \subfloat[][\centering $N=2$ components]{{\includegraphics[scale=0.29]{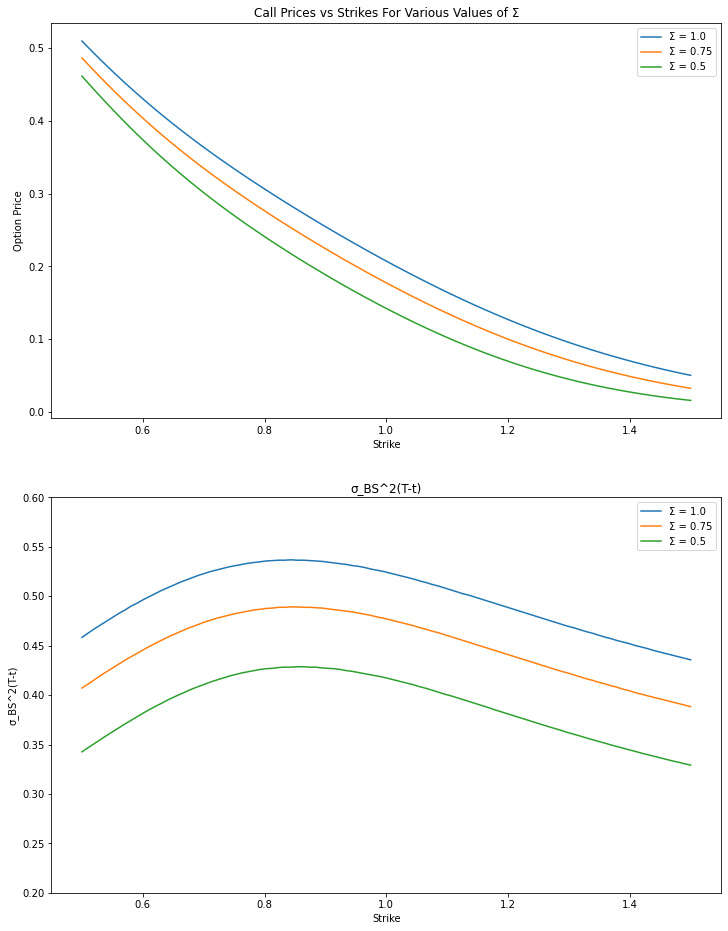}}}%
    \qquad
    \subfloat[][\centering $N=3$ components]{{\includegraphics[scale=0.29]{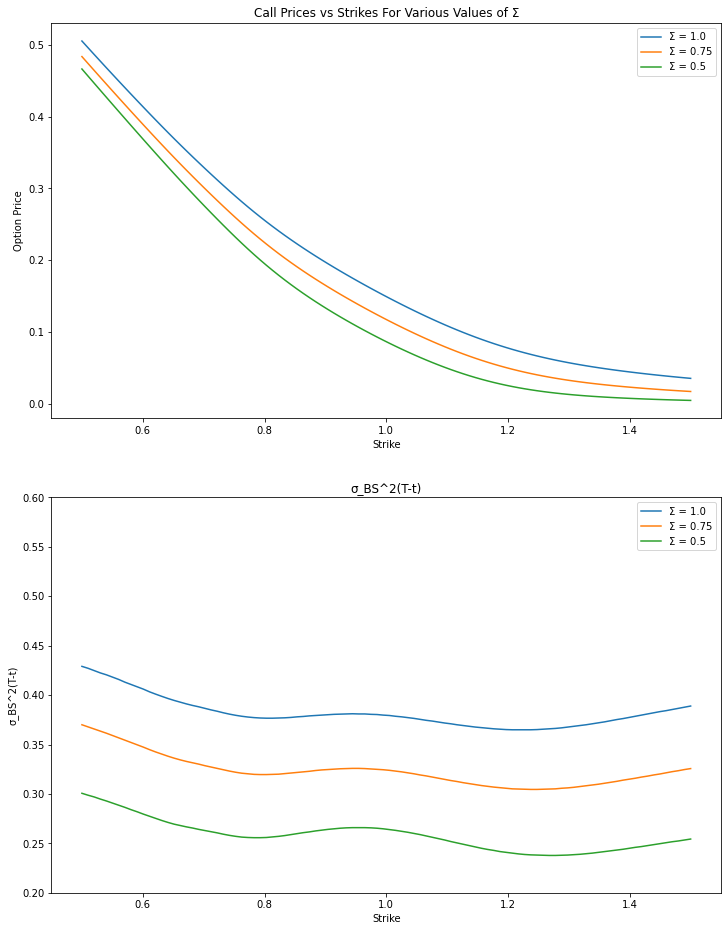}}}%
    \caption{$\sigma^2_{BS}(T-t)$ of Call Options with Gaussian Mixture as log price distribution}%
    \label{IBSRL Curves}%
\end{figure}

Since conditionally on $\cF^m_t$, $\xi^*_T$ is normal with mean $\xi^*_t$ and variance $\Sigma_t$, for a given option payoff $H$, we can price the option as in \eqref{optionprice}. Thus, we can compute numerically call option prices for any strike price $K$ given values $\xi$ of $\xi^*_t$ and $\Sigma$ of $\Sigma_t$. If the time to maturity $T-t$ is given, from the stock and option price, we can obtain the implied volatility $\sigma_{BS}$ predicted from the model by inverting the Black-Scholes formula.
In fact in this inversion, $\sigma_{BS}^2(T-t)$ is only a function of $\xi$ and $\Sigma$ and not a function of $T-t$. Thus, in order to eliminate a parameter, instead of $\sigma_{BS}$, we compute $\sigma_{BS}^2(T-t)$.
 In figure \ref{IBSRL Curves}, we plot the option prices and $\sigma_{BS}^2(T-t)$ 
 as a function of the strike $K$ for 3 different values of $\Sigma$ and the number of mixtures $N=2$ and $N=3$.  
 Corollary 4.1 of \cite{glasserman2021w} applies to our model and shows that $W$-shaped curves are not possible for $N=2$ components, as seen with the graph on the bottom left. On the other hand, for $N=3$ components, we obtain $W$-shaped curves. The implied volatilities and option prices are increasing in $\Sigma$

\subsection{Examples of volatility dynamics}\label{su.rough} 
Our goal in this
subsection is to establish a general sufficient condition for existence of exponential moments in Assumption \ref{asm:sigmanu}, \eqref{it:sigma}, and to apply it to two specific 
examples of commonly used volatility models, namely
the classical CIR model and its fractional counterpart. Without loss of generality, we assume that $T=1$ in throughout this subsection.

\subsubsection{A general sufficient condition}
Our sufficient condition, stated in Proposition \ref{pro:suff} below,  is based on 
the classical 
Garsia-Rodemich-Rumsey inquality (see \cite{garsia1970real}) whose
statement we reproduce here for the convenience of the reader:
\begin{theorem}[Garsia, Rodemich and Rumsey inequality] \label{grr}
    Suppose that $\Psi$ and $p$ are continuous
    and strictly increasing, $\Psi(0) = p(0) = 0$ and $\Psi(x) \to \infty$, as
    $x\to \infty$. For a continuous function $f:[0,1] \to \R$ let
    \begin{align*}
        c = \int_0^1 \int_0^1
        \Psi\Big( \frac{\abs{f(t) - f(s)}}{ p(\abs{t-s})}\Big)
        \, ds\, dt.
    \end{align*}
    If $c<\infty$ then
    \[ \abs{f(t) - f(s)} \leq 8
        \int_0^{t-s} \Psi^{-1}\Big( \frac{4c}{u^2} \Big)\, dp(u) \eforall\, 0\leq s < t
        \leq 1.\]
\end{theorem}
\begin{lemma} \label{exp-grr}
Let $X$ be a continuous process on $[0,1]$. For $\gamma_0 \in (0,1)$ and $M>0$ let
\begin{align} \label{F}
    F_M =\int_0^1 \int_0^1
    \exp{M\frac{\abs{X_t - X_s}}{ \abs{t-s}^{\gamma_0}}}
    \, dt\, ds.
\end{align}
If\, $\ee{F_M}<\infty$, for all $M>0$ then the pathwise $\gamma$-H\"
older seminorm
\[ \abs{X}_{\gamma} = \esssup_{0\leq s< t\leq 1}
    \frac{\abs{X_t - X_s}}{\abs{t-s}^{\gamma}}\]
admits all exponential moments, for
each $\gamma \in [0,\gamma_0)$.
\end{lemma}

\begin{proof}
It follows directly from Theorem \ref{grr} with  $p(u) = u^{{\gamma_0}}$  and $\Psi_M = \exp{M
x}-1$ that, for each $M>0$, with  $F_M$ is as in the statement, we have
    \begin{align*}
        \abs{X}_{\gamma} \leq L_M(4F_M) \ewhere L_M(x)=\sup_{r \in (0,1]} 8 r^{-\gamma} \int_0^{r}
        \Psi^{-1}_M\Big( \frac{x}{u^2} \Big) \, dp(u).
    \end{align*}
    For $x\geq 2$ and  $r \in (0,1)$ we have
    \begin{align*}
        M\int_0^r \Psi^{-1}_M\prn{\frac{x}{u^2}}\, d p(u) & 
        ={{\gamma_0}} \int_0^r \log\prn{1+\frac{x}{u^2}} u^{{{\gamma_0}}-1}\, du
        \leq  {{\gamma_0}}\int_0^r 
        \log\left(\frac{3}{2}\frac{x}{u^2}\right) 
        u^{{{\gamma_0}}-1}\, du  \leq L(r),
    \end{align*}
    where
    \begin{align*}
    L(r) = 2 {{\gamma_0}} \int_0^r \log\prn{\frac{x}{u^2}} u^{{{\gamma_0}}-1}\, du =
    \frac{2 r^{{\gamma_0}}}{{{\gamma_0}}} \prn{2+{{\gamma_0}} \log\prn{\frac{x}{r^2}}}. 
    \end{align*}
    We pick $\gamma \in [0,{\gamma_0})$ and observe that for $x$ large enough, 

    namely $x\geq x_0:= \exp{\tfrac{2 \gamma}{ {{\gamma_0}}({{\gamma_0}}-\gamma)}}$,
    the function 
    $$r \mapsto 8 r^{-\gamma} 
    L(r) = 8 r^{{{\gamma_0}}-\gamma} (2+{{\gamma_0}}\log(x/r^2))$$
    is non-decreasing on $(0,1]$. Therefore,
    \begin{align*}
        L_M(x) \leq  \frac{2 }{M}\left(\frac{2 }{{{\gamma_0}}}+  
        \log(\max(x,x_0,2))\right),
    \end{align*}
    so that, for each $k>0$ and each $M>0$, we have
    \[ \ee{ e^{k \abs{X}_{\gamma}}} \leq e^{4 k/(M {\gamma_0})}
        \ee{ e^{2 k/M \log(\max(4F_M,x_0,2))}}=
        e^{4 k/(M {\gamma_0})}\ee{ (\max(4F_M,x_0,2) )^{2 k/M} }.
    \]
    Given $k>0$, it remains to take $M = 2 k$ and use the assumption that $\ee{F_M}<\infty$.
\end{proof}
\begin{proposition}\label{pro:suff}
Suppose that there exists constants $\eps>0$ and ${\gamma_0}>0$ such that
\[ \sup_{s\ne t \in [0,1]} 
\ee{\exp{ \eps \frac{\abs{\sigma_t^2-\sigma^2_s}}{\abs{t-s}^{\gamma_0}}}} < \infty.\]
Then the $\gamma$-H\" older semi-norm of $\sigma$
    admits all exponential moments for any $\gamma \in [0,{\gamma_0}/2)$.
\end{proposition}
\begin{proof}
    Using the elementary fact that
    $\abs*{b-a} \leq \sqrt{\abs{b^2-a^2}}$ 
    for all $a,b \geq 0$, followed by
    Young's inequality, we obtain that for all $s\ne t \in [0,1]$,
    $M\geq 0$ and
    $\eps\in (0,1)$ we have
    \begin{align*}
        M \frac{\abs{\sigma_t - \sigma_s}}{\abs{t-s}^{{\gamma_0}/2}}  \leq
        M \sqrt{ \frac{\abs{\sigma^2_t - \sigma^2_s}}{\abs{t-s}^{{\gamma_0}}} }
        \leq \eps \frac{ \abs{\sigma^2_t - \sigma^2_s}}{\abs{t-s}^{\gamma_0}} + \frac{M^2}{4 \eps}
    \end{align*}
    We use Lemma \ref{sup-bnd-lem}
    to conclude that
    \begin{align} \label{pty}
        \ee{\exp{ M \frac{\abs{\sigma_t - \sigma_s}}{\abs{t-s}^{{\gamma_0}/2}}}} \leq {C \exp{ C \eps \kappa+\frac{M^2}{\e}}} < \infty,
    \end{align}
    Since the right-hand side of
    \eqref{pty} is independent of $s,t$, we conclude that, for each $M$,
    \[ \sup_{s\ne t \in [0,1]} \ee{ \exp{ M \frac{\abs{\sigma_t -
                        \sigma_s}}{\abs{t-s}^{{\gamma_0}/2}} }} < \infty.\]
    It remains to apply Lemma \ref{exp-grr}.
\end{proof}

\subsubsection{CIR-based volatility}
Let $(V_t)$ be a nonnegative CIR process on $[0,1]$ 
started at $x>0$ at
$t=0$. More precisely, we assume that $V$ admits the dynamics of the form
\begin{align}
  \label{CIR}
  V_t = x + \int_0^t (a - k V_s)\, ds + \int_0^t \eta \sqrt{V_t}\, dW_t,
\end{align}
where $a,k,\eta> 0$ are such that the Feller condition $2a \geq \eta^2$ is
satisfied; consequently, $V_t\geq 0$ for all $t \in [0,1]$, a.s. 
An explicit formula for the 
moment-generating function of $V_t$ (see \cite{Alf15}, Proposition
1.2.4, p.~7) is given by
\begin{align}\label{CIR-mgfk}
  \ee{ \exp{ u V_t}} = 
  F(t,u)^{2a/\eta^2} \Exp{ u \exp{-k t} F(t,u) x}, 
    \ewhere F(t,u) = \prn{1 - \frac{\eta^2}{2k}  \prn{1-e^{-kt}} u}^{-1},
\end{align}
for  $u< u_0= 2k \eta^{-2} (1-\exp{-k})^{-1}$.

Since $(1-e^{-1}) \leq \tfrac{1}{s}(1 - \exp{ - s}) \leq 1$, 
for all $s\in (0,1]$,
there exist constants $0<c_1< c_2$, that depend only on $a,k$ and $\eta$,
such that
\begin{align*}
  (1- c_1 \eps \sqrt{t})^{-1} \leq  F\prn{t, \frac{\eps}{\sqrt{t}}} \leq 
  \prn{1-c_2 \eps\sqrt{t}}^{-1} \efor t\in (0,1] \eand \eps\in \prn{-\infty, \tot \max(c_1,c_2)^{-1}}.
\end{align*} 
Therefore, there exists a constant $C$ such that 
 $\abs{F(t,\eps/\sqrt{t}) - 1} \leq C \sqrt{t}$ for 
 $t\in (0,1]$ and $\eps<\tot \max(c_1,c_2)^{-1}$.
Perhaps with a different constant $C$ we then also have
\begin{align}\label{F-est}
 \abs{\exp{-kt}F(t,\eps/\sqrt{t}) - 1} \leq C \sqrt{t}
\efor \eps<\tot \max(c_1,c_2)^{-1}.
\end{align}
This estimate implies, in particular, that $F(t,\eps/\sqrt{t})$ admits 
an upper bound, uniform in $t$ and all sufficiently small $\eps$. Therefore, for $\eps>0$ small enough, \eqref{CIR-mgfk} and \eqref{F-est}
imply that
\newcommand{\est}{\frac{\eps}{\sqrt{t}}}
\begin{align*}
  \ee{\exp{\eps \frac{\abs{V_t - x}}{\sqrt{t}} }}&\leq 
  \ee{\exp{\eps \frac{V_t - x}{\sqrt{t}} }}+
  \ee{\exp{-\eps \frac{V_t - x}{\sqrt{t}} }} \\
  &\hspace{-10ex}= F^{2a/\eta^2}\left(t,\est\right) \Bigg(
  \exp{ \est x \left(\exp{-k t} F\left(t,\est\right)  - 1\right) }\\ &+
   \exp{ \est x \left(1-\exp{-k t} F\left(t,-\est\right)\right) } \Bigg) \leq  
   C \exp{ C \eps x},
\end{align*}
for some constant $C$ which depends only on $a,k$ and $\eta$.
Since $V$ is a homogeneous Markov process, we deduce that
for all $0\leq s < t \leq 1$ and all $x\geq 0$, we have
\begin{align*}
  \ee{ \exp{ \eps \frac{\abs{V_t - V_s}}{\sqrt{t-s}}}} 
  \leq C \ee{ \exp{ C \eps V_{t-s}}}
\end{align*}
It follows directly from \eqref{CIR-mgfk} that, 
for small enough $\eps$, we
have $\sup_{\delta\in [0,1]} \ee{ \exp{ C \eps V_{\delta}}}<\infty$. Therefore, the conditions of Proposition \ref{pro:suff} are met, and, so 
for $0<\underline{\sigma}<\overline{\sigma}$, the process 
$\sigma_t = \underline{\sigma}+\min(\sqrt{V_t}, \overline{\sigma} )$ satisfies Assumption \ref{asm:sigmanu}, \eqref{it:sigma}.  

\subsubsection{The rough CIR model} Next, we consider
a class of models driven by Volterra
stochastic differential equations, including 
a truncation of the volatility process
in the rough Heston model,
as described in \cite{el2019characteristic,abi2019affine}. 
We mention also that in \cite{biagini2012insider} a Kyle's model 
where the noise trading is assumed to be a fractional process is studied.

We fix $H \in (0,1/2)$, set $\alpha = H+1/2$ and let 
\begin{align}\label{K-kernel}
K( t)= t^{\alpha-1}/\Gamma(\alpha)
\end{align}
be the $\alpha$-fractional kernel, where $\Gamma$ denotes the Gamma
function.  We choose two bounded continuous functions $b,s:\R\to\R$
and a constant $\kappa$ such that
\begin{align*}
    \abs{b(x)},s^2(x) \leq \kappa, \eforall  x\in\R.
\end{align*}
We also assume that the stochastic Volterra equation 
\begin{align}\label{eq:rheston}
    V_t = V_0 + \int_0^t K(t-u) b(V_u)\, du +
    \int_0^t K(t-u) s(V_u)\, dW_u,\ t\in [0,1],
\end{align}
admits a strong solution.
One sufficient condition is the Lipschitz continuity of 
the coefficients $b$ and $s$ (see
\cite{abi2019affine}, Theorem 3.3., p.~3167). 
A (financially) 
more relevant case occurs when  $b$ and $s$ define
\define{truncated Volterra square-root process}, i.e., when 
\begin{align}\label{trunc-CIR}
   V_t = V_0 + \int_0^t K(t-u) (b^0 - b^1 V_u)\, du 
   + \int_0^t K(t-u) \sqrt{A^1 \min(V_t, \oV)  }\, dW_u
\end{align}
where $b^0, b^1, A^1$ and $\oV$ are nonnegative constants. The existence,
uniqueness and positivity theory of \eqref{trunc-CIR}, parallels closely that 
of its version presented in Lemma 6.3, p.~3182 of \cite{abi2019affine}, so we
do not go into details here.   

Fix $0<\underline{\sigma}<\overline{\sigma}$. To show that the process $\sigma_t=\underline{\sigma}+\min(\sqrt{V_t}, \overline{\sigma} )$ satisfies
Assumption \ref{asm:sigmanu}, \eqref{it:sigma}, we employ
Proposition \ref{pro:suff} above to $t\mapsto \sqrt{V_t}$. Lemma \ref{sup-bnd-lem} below 
makes sure that its conditions are satisified.
\begin{lemma} \label{sup-bnd-lem}
    There exists a constant $C>0$,
    which depends only on $\alpha$ and $\kappa$ such that
    \begin{align} \label{est}
        \ee{\exp{ \eps \frac{\abs{V_t - V_s}}{\abs{t-s}^{H}}}} \leq
        {C \exp{ C \eps \kappa}} \eforall s\ne t \eand \eps \in [0,1).
    \end{align}
\end{lemma}
\begin{proof}
    Given $0\leq s < t \leq 1$, we decompose
    \[ V_t - V_s = A + M + D + N\]
    where
    \begin{align*}
        A & = \int_0^s (K(t-u) - K(s-u)) b(V_u)\, du   \\
        M & = \int_0^s (K(t-u) - K(s-u)) s(V_u)\, dW_u \\
        D & = \int_s^t K(t-u)  b(V_u)\, du             \\
        N & = \int_s^t K(t-u)  s(V_u)\, dW_u.
    \end{align*}
    For $r \geq 0$, the Cauchy-Schwarz inequality implies that
    \begin{align*}
        \ee{ \exp{ r\abs{V_t - V_s} }} & \leq
        \ee{ \exp{ r (\abs{A} + \abs{D})} \exp{  r \abs{M + N} } }                                                      \\
                                       & \leq \ee{ \exp{ 2r(\abs{A}+\abs{D}} }^{1/2} \ee{ \exp{ 2r \abs{M+N} } }^{1/2}.
    \end{align*}
    With $C$ denoting a generic constant, which may change from use to
    use and is allowed to depend only on
    $\alpha$ and $\kappa$, we have
    \begin{align*}
        \int_s^t K(t-u)\, du & = C  (t-s)^{\alpha} \eand
        0\leq \int_0^s K(s-u) - K(t-u)\, du {=C(s^\a-t^\a+(t-s)^\a)\leq} C  (t-s)^{\alpha}.
    \end{align*}
    Combining the equalities above with fact that $(t-s)^{\alpha} \leq (t-s)^{H}$,
    we obtain that
    \begin{align*}
        \abs{A} & \leq C \kappa \int_0^s (K(s-u) - K(t-u))\, du \leq C
        \kappa (t-s)^{\alpha} \leq C \kappa (t-s)^{H} \eand            \\
        \abs{D} & \leq C \kappa \int_s^t K(t-u)\, du
        \leq C \kappa (t-s)^{\alpha}
        \leq C \kappa (t-s)^{H}
    \end{align*}
    Consequently,
    \begin{align} \label{fv-est}
        \ee{\exp{2 r (\abs{A} + \abs{D})}} \leq { \exp{ C r \kappa (t-s)^H}}.
    \end{align}
    Turning to the $M+N$-term, we note that $M+N
        = \int_0^t H_u\, dW_u$ where
    \begin{align*}
        H_u & = \begin{cases}
                    (K(t-u) - K(s-u)) s(V_u) & u\in [0,s], \\
                    K(t-u)  s(V_u)           & u\in [s,t].
                \end{cases}
    \end{align*}
    Using the inequality
    $\ee{ \exp{ \pm  L_t}} \leq \ee{\exp{2 \ab{L}_t}}^{1/2}$, valid for any
    continuous local martingale $L$,
    we conclude that
    \begin{align*}
        \ee{ \exp{ 2r  \abs{M+N}}} \leq \ee{ \exp{ 2 r (M+N)}} + \ee{ \exp{ - 2 r (M+N)}}
        \leq 2 \ee{\exp{ 8 r^2 \int_0^t H_u^2\, du}}^{{1/2}}.
    \end{align*}
    For $0\leq u\leq s\leq t$, we have that
    \begin{align*}
        0 & \leq ((s-u)^{\a-1}-(t-u)^{\a-1} )^2 \leq (s-u)^{2\a-2}-2(s-u)^{\a-1}(t-u)^{\a-1} +(t-u)^{2\a-2} \\
          & \leq (s-u)^{2\a-2}-(t-u)^{2\a-2}.
    \end{align*}
    The (in)equalities
    \begin{align*}
        \int_s^t K(t-u)^2\, du & = C (t-s)^{2\alpha-1} \eand
        \int_0^s (K(t-u) - K(s-u))^2\, du \leq C  (t-s)^{2\alpha-1}
    \end{align*}
    further yield
    \begin{align*}
        \ee{ \exp{2r\abs{M+N}}} & \leq
        2 \ee{ \exp{ 8 r^2 \Big( \int_0^s (K(s-u) - K(t-u))^2 \eta(V_u)^2\, du
        + \int_s^t K(t-u)^2 (\eta(V_u))^2\, du \Big)}}^{{1/2}}                              \\
                                & \leq 2{ \exp{ C r^2 \kappa (t-s)^{2\alpha -1} }}^{{1/2}}.
    \end{align*}
    Since $(t-s)^{2\alpha - 1} = (t-s)^{2 H}$, this implies that
    \begin{align}\label{m-est}
        \ee{ \exp{ 2r|M+N|}} \leq 2{ \exp{ C \Big(r (t-s)^{H}\Big)^2 \kappa }}^{{1/2}}
    \end{align}
    If we replace $r$ by $\eps/ (t-s)^H$, combine the estimates \eqref{fv-est} and
    \eqref{m-est} above, and use that $\eps^2<\eps$ for $\eps \in [0,1)$,  we get
    \begin{align*}
        \ee{ \exp{ \eps \tfrac{|V_t - V_s|}{\abs{t-s}^H} }} & \leq
        \sqrt{2} { \exp{ C \eps\kappa}}^{1/2}
        { \exp{ C \eps^2 \kappa}}^{1/4}  \leq C {\exp{C \eps \kappa}}\qedhere
    \end{align*}
\end{proof}

\section{Proofs} \label{s.proofs}
We divide this section into two parts. In the first part we work towards
the proof of the Proposition \ref{G}, while he second one focuses on the
proof of the main Theorem \ref{thm:main}.  We refer the
reader to subsection \ref{su:building-blocks} above for all unexplained
notation. 

\subsection{Proof of Proposition \ref{G}} We start with a modest 
generalization of the standard
existence and comparison result for Lipschitz BSDE in a special case. 
\begin{lemma}\label{Lip-BSDE}\
\begin{enumerate}
\item (Existence) \label{it:exist}Suppose that the random field $f: [0,T]\times \Omega
\times \R\to \R$ is $\cF^W$-progressively measurable, $f(t,0) \in
\cS^{\infty}$,
\begin{align}\label{f-Lip}
    | f(t,y') - f(t, y)| \leq C |y'-y| \text{ for all $y,y'\in\R$ and all
        $t$, a.s.}
\end{align}
and that $b$ is a $\bmo$ process. Then the BSDE
\begin{align} \label{equ:Lip-BSDE}
    y_t = \int_t^T \Big(f(u,y_u) + z_u b_u\Big)\, du + \int_t^T z_u\, dW_u,
\end{align}
admits a solution $(y,z)\in \cS^{\infty} \times \bmo$.
\item (Comparison) \label{it:compar} Suppose that $b\in\bmo$,  $f^1$ and $f^2$ are
$\cF^W$-progressive random fields which satisfy \eqref{f-Lip} above  and
$f^1(t,y) \leq f^2(t,y)$ for all $t,y$, a.s. If $(y^1,z^1)$ and
$(y^2,z^2)$ in $\cS^{\infty} \times \bmo$ satisfy
\[ y^i = \int_t^T (f^i(u, y^i_u) + z^i_u\, b_u)\, du + \int_t^T z^i_u\,
dW_u,\] then $y^1_t \geq y^2_t$, for all $t$, a.s.  In particular, the
solution in 1.~above is unique.
    \end{enumerate}
\end{lemma}
\begin{proof}
(1)\   Let $\Gamma_t =\cE(\int_0^{t} b_u\, dW_u)$, and let $\E^{b}$
denote the expectation under the probability measure $\P^b$ defined by
$d\P^b = \Gamma_T\, d\P$.   We define  $\Phi:\cS^{\infty} \to
\cS^{\infty}$ by
\[ \Phi(y)_t = \eeup{b}{ \int_t^T f(u,y_u)\, du \giv  \sF^W_t }\] and note
that the Lipschitz property of $f$ implies that
\[ \|{\Phi(y^2)_t - \Phi(y^1)_t}\|_{L^{\infty}} \leq C \int_{t}^T
\|{y^2_u- y^1_u}\|_{L^{\infty}}\, du.\] From there, it follows
immediately that $\Phi$ is a contraction on $\cS^{\infty}$  equipped
with the (Banach) norm $\|y\|_{w} = \sup_t e^{ 2C t}
\|y_t\|_{L^{\infty}}$. The fixed point $\hat{y}$ of $\Phi$ has the
property that $\hat{y}_T=0$ and $\Gamma_t \hat{y}_t+\int_0^t \Gamma_u
f(u,\hat{y}_u)\, du$ is a $\P^b$-martingale satisfying
$$\sup_{t\in [0,T]}  |\Gamma_t \hat{y}_t|+\int_0^t |\Gamma_u
f(u,\hat{y}_u)|\, du\leq C \sup_{0\leq t\leq T}|\Gamma_t|.$$
Theorem 3.1., p.~57, in \cite{kazamaki2006continuous}
applied to $\Gamma$ and the martingale representation theorem imply that
\eqref{equ:Lip-BSDE} holds for some  progressive process $\hat z$ with
$\E^b\left[\sqrt{\int_0^T \hat{z}_u^2}\, du\right]<\infty$. Since $\hat
y_t$ and $f(t,\hat y_t)$ are both bounded processes, the process
$\int_0^t \hat z_u (-b_u\, du + dW_u)$ is a bounded $\P^{b}$-local
martingale, and, therefore, a $\P^{b}$-BMO martingale. Thanks to the
invariance of $\bmo$ spaces under equivalent measure changes (see
\cite[Theorem 3.3., p.~57]{kazamaki2006continuous}), $\hat z$ is a bmo
process with respect to $\P$, as well.

(2)\ This can be proved using the standard linearization method (see,
e.g., {the proof of} \cite[Theorem 4.4.1, p.~87]{zha17}) while keeping
in mind the fact that $\cE(\int_0^{\cdot} b_u\, dW_u)$ is a uniformly
integrable martingale as soon as $b$ is a $\bmo$-process (see
\cite[Theorem 2.3, p.~31]{kazamaki2006continuous}).
\end{proof}

\begin{proof}[Proof of Proposition \ref{G}]
    \newcommand{\ton}{\tfrac{1}{n}}
    \newcommand{\ty}{\tilde{y}}
    \newcommand{\tz}{\tilde{z}}
    The first step is to solve the auxiliary BSDE 
    \begin{align}\label{aux-BSDE}
        y_t = \int_t^T  \Big( \frac{\tilde J^2_u}{ 2 y_u} 
        - z_u \psi_u\Big) \, du + \int_t^T z_u\, dW_u,
    \end{align}
where $\tilde J_t = \hrho J_t$ and $\psi$ is given by $L = L_0
\sE(\int_0^{\cdot} \psi_t\, dW_t)$. 
    To do that, we pose a sequence of BSDEs
    \begin{align}\label{aux-BSDE-eps}
        y^{n}_t = \int_t^T \Big( \frac{\tilde J^2_u}{2(\ton \tilde J_0 + |y^{n}_u|)} - z^{n}_u \psi_u \Big)\, du +
        \int_t^T z^{n}_u\, dW_u,
    \end{align}
    each of which has a unique solution $(y^{n}, z^{n}) \in \cS^{\infty}
    \times \bmo$ by Lemma \ref{Lip-BSDE}, \eqref{it:exist} above. The second
    part of the same Lemma implies that
\[ n_1 \leq n_2 \implies y^{n_1}_t \leq y^{n_2}_t \eforall t\in [0,T],
\eas.\]
Assuming, without loss of generality, that we are working on the canonical
space $C[0,T]$, we define the adapted processes
    $$\tilde J^+(t,\omega) = \sup_{s\geq t,\tilde \omega}
        {|\tilde J_s(\omega\oplus_t \tilde \omega)|},\, \tilde J^-(t,\omega) = \inf_{s\geq t,\tilde \omega}
        {|\tilde J_s(\omega\oplus_t \tilde \omega)|}$$ and for fixed $t\in [0,T]$ consider two additional families of BSDEs
    \begin{align}\label{ypm}
        y^{t,\pm , n}_s = \int_s^T \Big( \frac{(\tilde J_t^\pm)^2  }{2( \ton \tilde J_0 +|y^{t,\pm , n}_u|)}
        - z^{t,\pm ,n}_u \psi_u \Big)\, du + \int_s^T z^{t,\pm , n}_u\, dW_u.
    \end{align}
    For each $n\in\N $, these equations admit  positive deterministic 
    (conditionally on $\cF^W_t$) solutions:
    \begin{align*}
        y^{t,\pm,n}_s &=  \sqrt{ \tfrac{\tilde J_0^2}{n^2}  + (\tilde J^\pm_t)^2(T-s)} -\frac{\tilde J_0}{n} , \ z^{t,\pm,n}_s=0.    \end{align*}
    Moreover, since  Lemma \ref{Lip-BSDE}, \eqref{it:compar} applies to these equations as well, we have
    \begin{align*}
        y^{n}_t \leq y^{t,+,n}_t \leq
        \tilde J^+_t \sqrt{ T-t}
    \end{align*}
    as well as
    \begin{align}\label{dominated}
        \ton \tilde J_0+y^{n}_t \geq \ton \tilde J_0  + y^{t,-,n}_t \geq   \tilde J^-_t\sqrt{T-t}.      .
    \end{align}
    Combining both inequalities, we obtain the following
    estimates:
    \begin{align}\label{y-order}
        \tilde J^-_t\sqrt{T-t} -\ton \tilde J_0 \leq y^{n}_t \leq
        \tilde J^+_t \sqrt{ T-t}
    \end{align}
    Let the process $y$ be defined by $y_t = \liminf_{n\to\infty}
        y^{n}_t$. It is progressively measurable and we have $y_t = \lim_n y^{n}_t$, a.s., for each
    $t$.
    Thanks to \eqref{y-order}  the sequence
    $(\tilde J_0/n + y^n_t)^{-1}$ is dominated by the integrable function $(T-t)^{-1/2}$
    (up to a multiplicative constant), so that the dominated convergence theorem
    can be applied to
    pass to the limit $n\to\infty$ on both sides of the equality
    $y^n_t = \E^{\psi}_t [ \int_t^T \tilde J^2_u/(2 \tilde J_0/n + 2y^{n}_u)\, du]$ to conclude that
    \[ y_t = \E^{\psi}_t\left[ \int_t^T \frac{  \tilde J^2_u}{2 y_u}\, du \right], \text{ a.s.,
            for each $t$.}
    \]
The fact the augmented filtration of $W^{\psi} = W - \int_0^{\cdot}
\psi_u\, du$ is Brownian
    allows us to show that $y$ admits a continuous modification, and, then,
    as in the proof of Lemma \ref{Lip-BSDE} above, that $(y,z)$ satisfy the BSDE
    \eqref{aux-BSDE} for some $z\in\bmo$.

    Passing to the limit in \eqref{y-order} we obtain
    \begin{align}\label{yJ-bnds}
        \tilde J^-_t\leq h_t \leq
        \tilde J^+_t
        \text{ where } h_t = \frac{y_t}{  \sqrt{T-t}}.
    \end{align}
    The two bounds above imply that $h$ is positive, bounded, bounded away from $0$ and $\lim_{t\to T}
        h_t = \tilde J_T$.
    Recalling that $|\tilde J|_\gamma$ denotes the $\gamma$-Holder norm of $\tilde J$,
    It\^ o's formula yields the following dynamics for $h$ on $[0,T)$
    \[ dh_t = \Big( a_t + K_t\, \psi_t \Big)\, dt - K_t\, dW_t,\]
    where $a_t = \tfrac{\tilde J_t+h_t}{2 h_t} \frac{h_t - \tilde J_t}{T-t} $ is  bounded by $\left(\frac{\tilde J_t}{2h_t}+\frac{1}{2}\right)|\tilde J|_\gamma |T-t|^{1-\gamma}$
    and $K_t=\tfrac{z_t}{ \sqrt{T-t}}=\frac{h_t z_t}{y_t}$.
    We apply It\^o's formula a second time to obtain
    \[ d\ln h_t = \left( \frac{a_t}{h_t} +\frac{1}{2}\psi^2_t - \frac{1}{2}\left(\frac{z_t}{y_t}-\psi_t\right)^2\right)\, dt - \frac{z_t}{y_t}\, dW_t\]
    which thanks to the identity $\sigma_t=L_t J_t=J_te^{\int_0^t \psi_s dW_s-\frac{1}{2}\int_0^t \psi_s^2 ds}$ leads to identity
    \begin{align}\label{eqbh}   \frac{h_t}{J_t e^{\int_0^t \frac{a_s}{h_s} \,ds } }
        \sigma_t = h_0 e^{\int_0^t \psi_s-\frac{z_s}{y_s}\, dW_s- \frac{1}{2}\int_0^t\left(\psi_s-\frac{z_s}{y_s}\right)^2ds }.
    \end{align}
    Assumption \ref{asm:sigmanu}, \eqref{it:sigma} 
    and the bounds of $a$ imply that
    the term $e^{\int_0^t \frac{a_s}{h_s} \,ds }$ admits all moments.
    The right hand side of \eqref{eqbh} is proportional to $1/\lambda_t$.
    In fact a direct computation shows that \eqref{eqbh} leads to
    \begin{align}
        \hrho^2 \sigma_t=\sqrt{T}h_0\sigma_0\gamma_t e^{\int_0^t \psi_s-\frac{z_s}{y_s}\, dW_s- \frac{1}{2}\int_0^t\left(\psi_s-\frac{z_s}{y_s}\right)^2ds }.
    \end{align}
    where $\gamma_t:=\tilde J_t \frac{ e^{\int_0^t \frac{a_s}{h_s} \,ds } }{h_t\sqrt{T}}$.
    We can compute
    \begin{align*}
        \int_0^T \gamma^2_t dt & =\int_0^T   \frac{\tilde J^2_t}{Th^2_t} e^{\int_0^t 2\frac{a_s}{h_s} \,ds } dt=\int_0^T   \frac{\tilde J^2_t}{Th^2_t} e^{\int_0^t \frac{1-\tilde J^2_s/h^2_s}{(T-s)}   \,ds } dt \\
                               & =\int_0^T   \frac{\tilde J^2_t}{h^2_t(T-t)} e^{-\int_0^t \frac{\tilde J^2_s}{(T-s)h^2_s}   \,ds } dt=1
    \end{align*}
    which is deterministic.

    As the last step in the existence proof, we define $G_t = y_t^2 L_t^2$ where
    $L_t = \cE(\int_0^{\cdot} \psi_u\, dW_u)$. A direct computation implies that
    $G$ satisfies the original BSDE \eqref{bsdeG} with
    \begin{align}\label{UG} {U_t} = {2G_t}\Big(\psi_t - \frac{z_t}{y_t}\Big) .
    \end{align}

    Since $\psi\in\bmo$, $y_t \geq C \sqrt{T-t}$ for some constant $C$ and
    $z_t/\sqrt{T-t}\in\bmo$, the equality \eqref{UG} above implies that
$\tfrac{U}{G} \in \bmo$. On the other hand, the relation \eqref{yJ-bnds}
implies that $G_t \leq \eta (T-t)$ for some strictly positive random variable
$\eta$. With $\sigma$ being continuous and bounded away from $0$, this
implies that $\int_0^T \tfrac{\sigma_s^2}{G_s}\, ds = \infty$, a.s. 

    \smallskip

    To prove uniqueness, let $(G^i,U^i)\in \cS^{+}_0 \times \cP^2$ for $i=1,2$ be two solutions of \eqref{bsdeG}. By a direct computation,
    $y^i=\frac{\sqrt{G^i}}{L}$ solves \eqref{aux-BSDE}. We define $\delta y = y^2-y^1$ and $\delta z =
        z^2-z^1$,  and observe that
    \[ d \delta y_t = r_t \delta y_t \, dt + \delta z_t ( dW_t
        + \psi\, dt) \text{ where } r_t = \frac{\tilde J^2_t }{2 y^1_t y^2_t}\]
    on $[0,T)$.
    Since $r_t \geq 0$, the process $R$ defined by $R_t = \exp{- \int_0^t r_u\,
            du}$ for $t<T$ is positive and bounded, and admits a limit as $t\to T$.
    It\^ o's formula implies that $R_t \delta y_t$ is a bounded local martingale
    on $[0,T)$ with $R_t \delta y_t \to 0$ as $t\to T$. It follows that $R_t
        \delta y_t$ is a uniformly integrable martingale with the last element $0$ so that $R_t \delta
        y_t = 0$, for all $t$. since $R_t>0$ for $t<T$, we conclude that
    $y^1_t=y^2_t$ for all $t$, a.s., and $z^1 = z^2$ $\text{Leb} \times \P$-a.e.
\end{proof}

\subsection{Proof of Theorem \ref{thm:main}} 
\label{su:proof}{
The proof starts with the introduction of several processes used in the construction of the novel state 
process $\xi^*$ of \eqref{def:state2}. 
After that, Lemma
\ref{nine-part} collects some of their essential properties and  establishes the existence and uniqueness of   $\xi^*$; it also 
features three other facts needed in the sequel.
The rest of this subsection lays out the details of the proof and is divided
into several subsections. We refer the reader to subsection
\ref{su:building-blocks} above for all unexplained notation, and note
that Assumption \ref{asm:sigmanu} is in force throughout.}


We start with the auxiliary
$\sF^{W,B}$-martingale $\hZ$ given by 
\begin{align}\label{hZ}
\hZ_t = \int_0^t \hrho \lambda_s \sigma_s\, dB_s,
\end{align}
which, by the Dambis-Dubins-Schwarz theorem, admits the representation 
\begin{align*}
	\hZ_t = \beta_{\ab*{\hZ}_t} \ewhere \ab*{\hZ}_t =
	\int_0^t \hrho^2 \lambda_s^2 \sigma_s^2\, ds = 1 - \Sigma_t \in \cF^W_t,
\end{align*}
and where $\beta$
 is an $\sF^{W,B}$-Brownian motion given by
\begin{align} \label{beta-Gamma}
	\beta_u = \hZ_{\Gamma_u} \ewhere \Gamma_u =
	\inf \{ t\geq 0\, : \, 1-\Sigma_t = u \},
\end{align}
which satisfy $\Gamma_{1-\Sigma_t}=t$ amd $1-\Sigma_{\Gamma_u}=u$ thanks to continuity and strict monotonicity of $\Sigma_T$.
With $\hxi$ defined on $[0,1]$ by
\begin{align}\label{SDE-hxi}
	\hxi_t =  t h^{-1}(\tv) + (1-t) \int_0^t \oo{1-s}\, d\beta_s,
	\efor   t\in [0,1) \eand \hxi_1 = h^{-1}(\tv),
\end{align}
we set
$$\xi^*_{t} = \hxi_{1-\Sigma_t}.$$

\begin{lemma}\label{nine-part}\ 
\begin{enumerate}
\item \label{it:h-int} 
For all $\xi \in \R$ and $u\in (0,1]$, we have
\begin{align*}
 \int \prn{\int_0^{\xi+\zeta} \abs{h(x)}\, dx} p(u,\zeta)\, d\zeta <\infty.
\end{align*}
    \item \label{it:beta-Brown} 
	The Brownian motion $\beta$ is independent of $W$ and $\tv$.
    \item \label{it:hxi-bridge}
	 Conditionally on $\tv$,  $\hxi$ is a Brownian bridge from  
    $0$ to $h^{-1}(\tv)$ and admits the following dynamics:
\begin{align}\label{hxi-dyn}
  d\hxi_t = \frac{h^{-1}(\tv) - \hxi_t}{1-t}\, dt + d\beta_t 
  \efor t\in [0,1), \ \lim_{t\to 1}\hxi_t = 
  \hxi_1= h^{-1}(\tv), \eas
\end{align}
\item  \label{it:hxi-Brown}
$\hxi$ is a Brownian motion independent of $W$.
\item \label{it:Rxi-sq-int} For $u \in [0,1]$ and $p\in [1,2]$ we have
\begin{align}\label{Rxi-sq-exp}
  \ee{ \abs{R_{\xi}(1-u, \hxi_u)}^p } \leq \ee{ \abs{\tv}^p}<\infty.
\end{align}
\item \label{it:Rc-finite} $\ee{R^c(0,\tv)}<\infty$. 
\item \label{it:xis-mart} 
The random variable $\xi^*_T$ is
$\sF^{\xi^*,W}_t$-conditionally normal with mean $\xi^*_t$ and variance
$\Sigma_t$. In particular, $\xi^*$ is an $\sF^{\xi^*,W}$-martingale.
\item \label{it:orthog}{The process $\xi^*$ is a 
continuous $\sF$-semimartingale on $[0,T]$, orthogonal to $W$. 
Moreover, it is the unique continuous process that 
satisfies \eqref{def:state2} on $[0,T)$.  }
\item \label{it:Yhs-dxis}
We have
\begin{align}\label{Yhs-dxis}
	d\xi^*_t = \ld_t\, d\hY^*_t,
\end{align}
and the pairs $(\xi^*,W)$, $(\hY^*,W)$ and $(Y,W)$ all generate the same
filtration. 
\end{enumerate}
\end{lemma}
\begin{proof}\ 
\begin{enumerate}
\item
Using the monotonicity of $h$ to 
justify the inequality 
$\abs{h(\xi+\sqrt{u} y)} \leq \abs{h(\xi)} + \abs{h(y+\xi)}$, for $u \in [0,1]$, 
and a simple change of variables, we obtain 
\begin{align*}
 \int &\prn{\int_0^{\xi+\zeta} \abs{h(x)}\, dx} p(u,\zeta)\, d\zeta 
 = \int \prn{ \int_{0}^{\xi+y \sqrt{u}} \abs{h(x)}\, dx} p(1,y)\, dy \leq \\
& \leq \int \max\Bprn{\abs{h(\xi+\sqrt{u} y)}, \abs{h(0)}} \abs{\xi+y\sqrt{u}} p(1,y)\, dy\\
& \leq \int \Bprn{\abs{h(0)} + \abs{h(\xi)} + \abs{h(\xi+ y)}} \abs{\xi+y\sqrt{u}} p(1,y)\, dy\\
&\leq C  + C \int \abs{h(r)}(1+\abs{r}) p(1,r-\xi)\, dr
= C + C \int \abs{h(r)} (1+\abs{r}) e^{-\xi^2/2} e^{\xi r} p(1,r)\, dr
\end{align*}
where $C$ is a finite constant which depends on $h$ and $\xi$. 
The last integral is finite by the Cauchy-Schwarz inequality because
both $h(r)$ and $(1+\abs{r}) e^{\xi r}$ are square-integrable with
respect to the Gaussian measure with density $p(1,r)$. 

\item Since $\hZ$ is a Brownian motion independent of $W$ and $\tv$,
and $\Gamma$
is a time change with respect to $\sF^W$, the process $\beta$ is a
continuous martingale, conditionally on $W$. Its quadratic variation
is Brownian, so, by L\'evy's criterion, $\beta$ is a Brownian motion
independent of $W$ and $\tv$.

\item By its definition, $\beta$ is independent of $\tv$, so $\hxi$ is a
Gaussian process conditionally on $\tv$. Moreover, its  conditional mean
and covariance functions, namely $t h^{-1}(\tv)$ and $\min(s,t) - st$,
match those of the Brownian bridge from $0$ to $h^{-1}(\tv)$. The
dynamics \eqref{hxi-dyn} is a direct consequence of \itos{} formula. 

\item The process $\hxi$ is $\sF^{\beta,\tv}$-adapted and both $\beta$ and
$\tv$ are independent of $W$, as established in \eqref{it:beta-Brown}
above, so $\hxi$ is independent of $\sF^W$. To see that it is a Brownian
motion, it is enough to observe that it is a Brownian bridge from $0$ to
an independent standard normal (or simply compute its mean and covariance
functions as above). 
\item Since $R_{\xi}(1-\cdot,\cdot)$ is a 
space-time harmonic function and $\hxi$ is a
Brownian motion, $V_u = R_{\xi}(1-u,\hxi_u)$ is a martingale with the terminal
condition $R_{\xi}(0,\hxi_1) = \tv$ and the inequality \eqref{Rxi-sq-exp} is
a direct consequence of Jensen's inequality.
\item 
Using the fact that the supremum in 
	\begin{align*}
R^c(0,\tv) = \sup_{\xi\in\R}\Bprn{ \xi \tv - \int_0^{\xi} h(x)\, dx}
\end{align*}
is attained at $\xi = h^{-1}(\tv)$, we obtain
\begin{align*}
	R^c(0,\tv) & =
	\abs{\tv h^{-1}(\tv) - \int_0^{h^{-1}(\tv)} h(x)\, dx}  =
	\abs{  \int_0^{h^{-1}(\tv)} (\tv - h(x))\, dx }\\
	&\leq \abs{h^{-1}(\tv)} \max_{r \in [0,1]} \abs*[\Big]{\tv - h(r h^{-1}(\tv))}
	\leq \abs{h^{-1}(\tv)} \Bprn{ \abs{\tv} + \abs{h(0)}},
\end{align*}
where the last inequality follows from the monotonicity of $h$.
The random variable $h^{-1}(\tv)$ is normally distributed and therefore
in $\ltwo$, and so, 
by Assumption \ref{asm:sigmanu}, \eqref{it:tv}, both
$h(0) h^{-1}(\tv)$ and $ h^{-1}(\tv) \tv$ belong to $\lone$. 
\item $\hxi$ is a Brownian motion independent of $W$ and $\Sigma$ is
$\sF^W$-adapted. Therefore, conditionally on $\sF^W_T$, $\xi^*_t =
\hxi_{1-\Sigma_t}$ is a centered Gaussian process with  independent
increments and 
deterministic variance function
$1-\Sigma_t$. Therefore, conditionally on $\sF^W_T \vee \sF^{\xi^*}_t$, 
the random variable $\xi^*_T$ is normally distributed with mean
$\xi^*_t$ and variance $\Sigma_t$. The statement now follows by
further conditioning on $\sF^W_t$. 

\item {
To show that $\xi^*$ solves \eqref{def:state2}, we use
\eqref{hxi-dyn} to get the following representation
\[  \xi^*_t = 
\int_0^{1-\Sigma_t} \frac{h^{-1}(\tv) - 
\hxi_u}{1-u}\, du + \beta_{1-\Sigma_t} = 
\int_0^{t} \frac{h^{-1}(\tv) - 
\xi^*_{s}}{\Sigma_s}\, \hrho^2 \sigma^2_s \ld^2_s\, ds, 
+ \hZ_t, \ t\in [0,T) 
\]
where we used the change of variable $s = \Sigma^{-1}_{1-u}$.
Since $d\hZ = \hrho \ld \sigma\, dB$, $\xi^*$ 
indeed satisfies \eqref{def:state2}.

For uniqueness, it suffices to note that the equation for the 
difference of two solutions a linear ODE with random coefficients. 
These coefficients are bounded a.s., on $[0,t]$ for each $t<T$, 
and our claim follows from Gronwall's inequality.

By \eqref{hxi-dyn} above, $\hxi$ is a Brownian bridge, conditionally on $\tv$. 
Therefore, we have
\[  \int_0^1 \frac{\abs*{ h^{-1}(\tv) - \hxi_u}}{1-u}\, du<\infty, \eas\]
The same change of variable as above, 
namely $s=\Sigma^{-1}_{1-u}$, allows us to conclude
that \eqref{def:state2} provides
an $\sF$-semimartingale decomposition of $\xi^*$ on the entire $[0,T]$.  
The orthogonality with $W$ is the direct consequence of 
the fact that $W$ and $B$ are orthogonal. 

}

\item The identity \eqref{Yhs-dxis} follows directly from \eqref{def:state2}. 
Since $\ld$ is bounded and bounded away from $0$,
$(\xi^*,W)$ and $(\hY^*,W)$ generate the same filtration. The pairs
$(\hY^*,W)$ and $(Y,W)$ generate the same filtratoin because the
difference $Y^* - \hY^*$ is $\sF^W$-adapted.  \qedhere
\end{enumerate}
\end{proof}
\subsubsection{An expression for $\Pi(X,\fP^*(X+Z))$}
Let $X$ be a trading strategy, $Y = X + Z$, $\hY = X - \int_0^{\cdot} \rho\sigma_s\,
	dW_s$,  $\xi = \fxi^*(Y)$, and $P =\fP^*(Y)$, with the functionals 
$\fxi^*$ and $\fP^*$ defined in \eqref{def:state} and \eqref{def-P} above. 
We also introduce the following shortcuts for a generic function $F$:
\begin{gather*}
	(\Delta F)_t = F(\Sigma_t,\xi_t) - F(\Sigma_t, \xi_{t-}),\eand
	(\Delta^{2,\pm } F)_t = (\Delta F)_t -
	F_\xi(\Sigma_t,\xi_{t\pm})\Delta \xi_t.
\end{gather*}
Thanks to It\^ o's lemma,  we have the
following expressions for the dynamics of $P$ and $[P,X]$:
\begin{align*}
	dP_t     & =-\hrho^2\lambda^2_t \sigma_t^2 R_{\xi t}
	(\Sigma_t,\xi_{t-})\, dt
	+R_{\xi\xi}(\Sigma_t,\xi_{t-})\, d\xi_t
	+\frac{1}{2}R_{\xi\xi\xi}(\Sigma_t,\xi_{t-})
	d[\xi, \xi]^c_t  + (\Delta^{2,-} R_{\xi})_t, \\
	d[P,X]_t & =  \ld_t R_{\xi\xi}(\Sigma_t,\xi_{t-} )\Bprn{
		 d[X,X]^c +\hrho\sigma_t\, d[X,B]^c}+(\Delta R_{\xi})_t \Delta X_t.
\end{align*}
Since $(\Delta^{2-} R)_t + (\Delta R_{\xi})_t \Delta \xi_t =
	(\Delta^{2+} R)_t$, applying Ito's formula to $\frac{\tv\xi_t-R(\Sigma_t,\xi_t)}{\lambda_t}$ and rearranging terms, 
it follows that
\begin{align}\label{Pi-form}
	\Pi(X,P)_t = \int_0^t (\tv-P_{s-}) dX_s-[ P,X]_t = (I) + (II) + (III) + (IV),
\end{align}
where
\begin{equation}\label{1234}
    \begin{split}
	(I)   & =\frac{\tv\xi_t-R(\Sigma_t,\xi_t)}{\lambda_t} + \frac{R(1,0)}{\lambda_0}, \\
	(II)  & =  -\int_0^t\lambda_s\frac{R_{\xi\xi}(\Sigma_s,\xi_{s-})}{2}\, d[X]^c
	+ \sum_{s\leq t} (\Delta^{2,+} R)_s,                                               \\
	(III) & = \int_0^t(R(\Sigma_s,\xi_{s-})-\tv\xi_{s-})\, d\frac{1}{\lambda_s} \eand      \\
	(IV)  & = \hrho\int_0^t (P_{s-}-\tv) \sigma_s dB_s.
    \end{split}
\end{equation}

\subsubsection{An upper bound for $\fP^*$-admissible strategies}
Suppose now that $X$ is $\fP^*$-admissible and let $P = \fP^*(X+Z)$. 
Moreover, let $(\tau_n)$ be a common reducing
sequence for the $(\sF_t)$-local martingales in $(III)$ and $(IV)$ of
\eqref{1234}.  By the convexity of $R(\Sigma_t,\cdot)$, we have
$\Delta^{2,+}R_t \leq 0 \leq \Delta^{2,-}R_t$. Therefore, part $(II)$ is 
non-positive for each $t$, and the admissibility of $X$ implies, via 
Fatou's lemma, that
\begin{multline}\label{ie1}
	\ee{ \Pi(X,P)_T  \giv \tv }    =
	\ee{ \liminf_{n} \Pi(X,P)_{\tau_n} \giv \tv } 
	                   \leq \oo{\ld_0} R(1,0) + \liminf_n
	\ee{ \oo{\ld_{\tau_n}}
	\Bprn{\tv \xi_{\tau_n} - R(\Sigma_{\tau_n}, \xi_{\tau_n})}\giv \tv }
\end{multline}
For all for all $\xi$ and $u$, we have
\begin{align*}
    \xi \tv - R(u, \xi) & \leq
    \sup_{\xi} \Bprn{ 
        \xi \tv - \int R(0,\xi+\zeta)p(u,\zeta)\, d\zeta 
        }\\
        &=
    \sup_{\xi} \Bprn{ 
            \int \bprn{ (\xi+\zeta) \tv - R(0,\xi+\zeta)}p(u,\zeta)\, d\zeta 
        }\\ &
    \leq  \int \sup_{\xi} 
            \bprn{ (\xi+\zeta) \tv - R(0,\xi+\zeta)}p(u,\zeta)\, d\zeta 
            = R^c(0,\tv).
\end{align*}
Since $0\leq R^c(0,\tv)$, by the $\sF$-martingale property of $\ld$, we have
    \begin{align}\label{upper-bound}
      \ee{ \Pi(X,P)_T  \giv \tv} \leq \oo{\ld_0} R(1,0) + 
      \liminf_n \ee{ \oo{\ld_{\tau_n}} R^c(0,\tv) \giv \tv } 
      = \frac{R(1,0) + R^c(0,\tv)}{\ld_0} 
    \end{align}

\subsubsection{$\fP^*$-admissibility of $X^*$}
The chain rule implies that
\begin{align} \label{dxi}
	d\xi^*_t = \frac{\hxi_1 - \hxi_{1-\Sigma_t}}{\Sigma_t}
	\, d(1-\Sigma_t)+ d\hZ_t, 
\end{align}
as well as
\begin{align}\label{Xs}
	X^*_t = 
	\int_0^{1-\Sigma_t} \oo{\ld_{\Gamma_u}}   \frac{ \hxi_1 - \hxi_u}{1-u}\, du.
\end{align}
We define $P^*_t = \fP^*(X^*+Z)_t$ and use the identities \eqref{dxi} and
\eqref{Xs}, together
with \eqref{Pi} above, to obtain the following expression for $\Pi(X^*,P^*)_t$:
\begin{equation} \label{Pis}
    \begin{split}
        \Pi(X^*,P^*)_t &= \int_0^t \oo{\lambda_s} 
        \Bprn{\tv - R_{\xi}(\Sigma_s, \xi^*_s)}
        \Bprn{h^{-1}(\tv) - \xi^*_s} \, \frac{d (1-\Sigma_s)}{\Sigma_s},\\
        &=
        \int_0^{1-\Sigma_t} \frac{1}{\ld_{\Gamma_u}} 
        \Bprn{\tv - R_{\xi}(1-u, \hxi_u)}
        \frac{\hxi_1 - \hxi_u}{1-u}\, du.
    \end{split}
\end{equation}
Therefore,
\begin{align*}
    \sup_{t\in [0,T]} \abs*[\Big]{\Pi(X^*,P^*)_t} \leq
    \int_0^1 \oo{\ld_{\Gamma_u}} \abs{h(\hxi_1)}  \tfrac{\abs{\hxi_1 - \hxi_u}}{1-u}\, du
    + \int_0^1 \oo{\ld_{\Gamma_u}} \abs{R_{\xi}(1-u, \hxi_u)}
    \tfrac{\abs{\hxi_1 - \hxi_u}}{1-u}\, du.
\end{align*}
Since $\tfrac{1}{\ld_{\Gamma_u}}$ is a martingale in its own filtration
and measurable with respect to $\sF^W_T$, and $\sF^W$ and $\hxi$ are
independent by Lemma \ref{su:building-blocks}, (3), we have
\begin{align} \label{ep}
    \ee{\sup_{t\in [0,T]} \abs*[\Big]{\Pi(X^*,P^*)_t}} & \leq
    \oo{\ld_0} \ee{ \int_0^1 \abs*{h(\hxi_1)}
    \tfrac{\abs{\hxi_1 - \hxi_u}}{1-u}\, du} +                \\ & \notag \quad +
    \oo{\ld_0}
    \ee{\int_0^1  \abs{R_{\xi}(1-u, \hxi_u)} \tfrac{\abs{\hxi_1 - \hxi_u}}{1-u}\, du}.
\end{align}
{By Lemma \ref{nine-part}, \eqref{it:hxi-Brown},
$\hxi$ is a  Brownian motion and
$\hxi_1 - \hxi_u$ is independent of $\hxi_u$. Thus, the finiteness of the
expectation on the LHS of \eqref{ep} above boils down to the finiteness
of 
\begin{align*}
&\ee{ \int_0^1 \abs*{h(\hxi_1)}
    \tfrac{\abs{\hxi_1 - \hxi_u}}{1-u}\, du}\leq
     \ee{\abs*{\tilde v}}\int_0^1 
    \tfrac{\ee{\abs{\hxi_1 - \hxi_u}}}{1-u}\, du <\infty \mbox{ and}\\
    &\int_0^1\ee{  \abs{R_{\xi}(1-u, \hxi_u)}}\ee{ \tfrac{\abs{\hxi_1 - \hxi_u}}{1-u}}\, du \leq\ee{\abs*{\hxi_1}}  \int_0^1 \frac{\ee{\abs{R_{\xi}(1-u, \hxi_u)}}}{\sqrt{1-u}}\, du.
\end{align*}
By the definition of $R$ we have the stochastic representation 
$R_\xi(1-u,\hxi_u)=\ee{h(\hxi_1)\giv\hxi_u}$. Thus, we have the finiteness by 
\begin{align*}
    \int_0^1 \frac{\ee{\abs{R_{\xi}(1-u, \hxi_u)}}}{\sqrt{1-u}}\, du\leq   \int_0^1 \frac{\ee{\abs{\ee{h(\hxi_1)\giv\hxi_u}}}}{\sqrt{1-u}}\, du\leq \int_0^1\frac{\ee{\abs{\tv}}}{\sqrt{1-u}}\, du <\infty.
\end{align*}}
We can, therefore, conclude that $X^*$ is $\fP^*$-admissible.

\subsubsection{$P^*$-optimality of $X^*$}
By \eqref{hxi-dyn} and the space-time harmonicity of $R(1-\cdot,\cdot)$,
we have
\begin{multline}
\int_0^{T}  \prn{\tv - R_{\xi}(1-u, \hxi_{u}) }\, 
    \frac{\hxi_1 - \hxi_{u}}{1-u} \, du   = \\
    = \int_0^{T}  \prn{\tv - R_{\xi}(1-u, \hxi_{u}) }\, d\hxi_u 
- \tv \beta_{ T}  + \int_0^{ T} R_{\xi}(1-u, \hxi_{u})\, d\beta_u\\
     = R^c(0,\tv) + R(1,0) - \tv \beta_T 
    + \int_0^T R_{\xi}(1-u, \hxi_u)\, d\beta_u.
    \label{ie2}
\end{multline}
By Lemma \ref{nine-part}, \eqref{it:Rxi-sq-int}, the $d\beta$-integral
in the last line of \eqref{ie2} is a martingale.  Since $\ld$, $\tv$ and
$\beta$ are independent of each other, we have
\begin{align*}
  \ee{ \oo{\ld_T} \prn{ \beta_T\tv - 
    \int_0^T R_{\xi}(1-u,\hxi_u)\, d\beta_u } \giv \tv}=0.
\end{align*}
Therefore, by \eqref{Pis} and the martingale property of $\oo{\ld_{\Gamma}}$,
and, then, by \eqref{ie2}, 
we have 
\begin{align*}
 \ee{ \Pi(X^*, P^*)_T \giv \tv} 
 &= \ee{ \int_0^T \oo{\ld_{\Gamma_u}} 
 (\tv - R_{\xi}(1-u, \hxi_u))
			\frac{\hxi_1 - \hxi_u}{1-u}\, du \giv \tv}\\
&=\ee{ \oo{\ld_T} 
\int_0^T \prn{\tv - R_{\xi}(1-u, \hxi_u)}
    \frac{\hxi_1 - \hxi_u}{1-u}\, du \giv \tv}\\
&=\ee{ \oo{\ld_T} \prn{R^c(0,\tv) + R(1,0)} \giv \tv}
=\frac{R^c(0,\tv) + R(1,0)}{\ld_0}
\end{align*}
Therefore, 
since it attains the upper bound \eqref{upper-bound},
$X^*$ is an optimal strategy for the insider. 

\subsubsection{Rationality of $\fP^*(Y^*)$}
To show that $P^*$ is a rational pricing rule we use
\eqref{Yhs-dxis} of Lemma \ref{nine-part}
to conclude that $[\hY^*,W]=0$. Therefore, 
\begin{align*}
	\fxi^*(Y^*) = \xi^* \eand \fP^*(Y^*)_t = R_{\xi}(\Sigma_t, \xi^*_t).
\end{align*}
which reveals that  $\fP^*(Y^*)$ is a time-changed (by $1-\Sigma_t$)
version of the martingale $R_{\xi}(1-u, \hxi_u)$. Hence $\fP^*(Y^*)$ is a
martingale itself with respect to $\sF^{X^*,W} = \sF^{\xi^*,W}= \sF^{m*}$, and the
rationality condition \eqref{rational} follows from the fact that
\begin{align*}
\fP^*(Y^*)_T = R_{\xi}(0, \xi^*_T) = h(h^{-1}(\tv)) = \tv. 
\end{align*}

\bibliographystyle{apalike}
\bibliography{ref.bib}

\end{document}